\documentclass[conference]{IEEEtran}
\IEEEoverridecommandlockouts
\usepackage{cite}
\usepackage{graphicx}
\usepackage{balance}
\usepackage{soul}
\usepackage{listings}
\usepackage{url}
\usepackage{amsmath}
\usepackage{amssymb}
\usepackage{array}
\usepackage{tabu}
\usepackage{color}
\usepackage{xspace}
\usepackage{caption}
\usepackage{subfig}
\usepackage{csvsimple}
\usepackage{algorithm}
\usepackage[noend]{algpseudocode}
\usepackage{adjustbox}
\usepackage{enumitem}
\usepackage{listings}
\usepackage{wrapfig}
\usepackage{csvsimple}
\usepackage{booktabs}
\usepackage{amsthm}
\usepackage[group-separator={,}]{siunitx}

\begin{document}

\title{Memory-Efficient Group-by Aggregates over Multi-Way Joins\\
}

\newcommand{\graphgen}{{\sc GraphGen}\xspace}
\mathchardef\mhyphen="2D 

\newcommand{\gtuple}[2]{$\langle$#1$,$ #2$\rangle$}
\newcommand{\tuple}[2]{$\big($#1$,$ #2$\big)$}
\newcommand{\kv}[2]{$\big[$#1$:$ #2$\big]$}

\newcommand{\amolnote}[1]{\textcolor{red}{\bf #1}}
\newcommand{\kostasnote}[1]{\textcolor{blue}{\bf #1}}
\newcommand{\topic}[1]{\vspace{-3.5pt}\smallskip \noindent{\bf #1:}}
\newcommand{\topicul}[1]{\smallskip \smallskip \noindent\underline{\bf #1:}}

\newcommand{\calV}{\mathcal{V}}
\newcommand{\calR}{\mathcal{R}}

\newcommand{\joinAgg}{{\sc Join-Agg}\xspace}

\theoremstyle{definition}
\newtheorem{definition}{Definition}[section]
\newtheorem{exmp}{Example}[section]

\newtheorem{axiom}{Axiom}[section]


\let\proof\relax
\let\endproof\relax

\author{\IEEEauthorblockN{Konstantinos Xirogiannopoulos}
\IEEEauthorblockA{\textit{Department of Computer Science} \\
\textit{University of Maryland}\\
College Park, USA\\
kostasx@cs.umd.edu}
\and
\IEEEauthorblockN{Amol Deshpande}
\IEEEauthorblockA{\textit{Department of Computer Science} \\
\textit{University of Maryland}\\
College Park, USA\\
amol@cs.umd.edu}
}

\maketitle

\begin{abstract}
  Aggregate computation in relational databases has conventionally been done using the standard unary aggregation and binary join operators. These operators implement the traditional model of computing joins between
  relations two at a time,  materializing the intermediate results, and then proceeding to do either sort-based or hash-based aggregate computation to derive the final result.
  This approach, however, can be dramatically sub-optimal in the case of {\em low-selectivity} joins, and often ends up generating large intermediate results even if the relations involved as well as the final
  result sets themselves are quite small. Moreover, many of the intermediate results may never be a part of the final result. 

  In this paper we propose a novel aggregate query processing technique that leverages \textit{graph data structures} towards efficiently answering aggregate queries over joins, \textit{without
      materializing} intermediate join results. We wrap this technique inside a \textit{multi-way} composite database operator called \joinAgg that combines \textit{Join} and \textit{Aggregation}. We
      provide a general framework for executing aggregation queries over arbitrary acyclic joins, involving any number of group-by attributes from any relation. We also present a thorough experimental evaluation on both real world and synthetic datasets. Our
      experiments show that our operators can achieve orders of magnitude lower \textit{query times} and \textit{memory requirements} compared to the traditional approach, even when implemented \textit{outside} of the database system.
\end{abstract}


\section{Introduction}
\label{sec:intro}
Traditionally, aggregate processing over conjunctive queries in RDBMSs has been done through the use of simple binary operators for executing joins, followed by a (typically separate) unary aggregation operator. The
simplicity of these operators has proven invaluable throughout the development of modern RDBMSs. Each simple operator enables the optimization of a very specific operation with a concise set of
parameters, inputs and outputs. This enabled simpler query optimization, since it is easier to create good cost models for simple operators than for complex ones.


Simplicity however often comes at the cost of performance. It is known that binary operators can lead to sub-optimal performance regardless of the query plan
used~\cite{avnur2000eddies,deshpande2004initial}. The main drawback with binary join operators in RDBMSs specifically, is the generation of intermediate join results, potentially with
materialization at the granularity of every join. Each individual join within a query plan may output an increasingly larger number of tuples, making the latter intermediate results unwieldy, especially as
they start becoming larger than memory. Pipelining operators help in some cases, but enumeration of the full intermediate result set though joining every single tuple is necessary and cannot be
easily avoided (e.g., bloom filters can help in some cases to filter out results, but break the classical model as well). These issues become significantly more pronounced in analytics settings where the
joins are often done on non-key attributes to derive higher-level insights (see examples below).
%

This has led to an increasing interest in the idea of \textit{multi-way} database operators. 
Eddies was one example of such an operator~\cite{avnur2000eddies}, where the benefits of combining multiple operators into one came from the ability to choose different execution paths for different tuples.
More recently, breakthroughs \textit{in worst-case optimal} join
algorithms~\cite{ngo2012worst} have shown that one can put tight bounds on the maximum possible number of tuples generated by a query, and then develop algorithms whose runtime guarantees match those
worst-case bounds. These breakthroughs have led to a variety of different query operators that take a \textit{multi-way} join approach over the traditional binary operator. The benefits seen by many of
these proposed operators typically come from the fact that the operator takes multiple relations that are part of a large conjunctive query into account \textit{simultaneously}.
This allows for avoiding the materialization of large intermediate results~\cite{veldhuizen2012leapfrog}, enables pruning out various portions of the computation based on complex
conditions~\cite{walenz2017optimizing}, or allows for exploiting more parallelism and fast set intersections toward the join result~\cite{aberger2015emptyheaded}.

In this paper, we focus on another very common combination of operators, namely a series of joins followed by a group-by aggregate.
The data graph paradigm proposed here is reminiscent of {\em factorized representation of conjunctive query results}, by Olteanu et al.~\cite{olteanu2012factorised}, and
the idea of a \textit{tuple hypergraph} that can cover all tuples in a query result~\cite{kara2018covers}. All of these provide compact representations of the
underlying join result, especially in presence of low-selectivity joins, with minor differences because of the specific goals behind their genesis. Our key contributions
here are a novel way to use such a structure for computing \textit{group by aggregates} efficiently over complex acyclic joins.


Several different works have considered the problem of executing group by aggregate queries against a factorized representation of a conjunctive
query~\cite{bakibayev2013aggregation,khamis2018ac,khamis2018functional,khamis2019boolean,schleich2019layered,schleich2016learning,olteanu2016factorized}. The key guarantees like constant-delay enumeration, however, do
not extend to the kind of group by queries we focus on in this work, e.g., the ``branching'' query $R_1(g_1,j), R_2(j,b), R_3(b,g_  3), R_4(b, g_2)$. Because all of $g_1,
g_2, g_3$ (group by attributes) need to be present in the output, either (a) one of the other attributes needs to be eliminated (which requires generation of a large
intermediate result), or (b) we have to iterate over all combinations of values for $g_1, g_2, g_3$ and compute the aggregate value for each combination (which can be
prohibitively expensive if either the sizes or the number of group by attributes is large). Our work here, thus, can be seen as exploring an alternative approach to computing
aggregates over the factorized representation.

As we discussed earlier, recent work on worst-case optimal joins~\cite{ngo2012worst,DBLP:conf/icdt/Veldhuizen14,koutris2016worst,ngo2014skew} shows how to avoid large intermediate results during execution of multi-way join queries;
Joglekar et al.,~\cite{joglekar2015aggregations,joglekar2016ajar} discuss how to generalize that to aggregate queries. Their approach is largely complementary to ~\cite{bakibayev2013aggregation}, as well as our line of work.
Recent work on FAQ~\cite{abo2016faq} proposed a generalized way of viewing a very common type of aggregation query called a Functional Aggregate Query which they see
parallels in multiple scenarios other than databases, e.g., matrix multiplication, probabilistic graphical models, and logic. For acyclic join queries, those approaches
effectively reduce to pushing aggregates below joins in the query plans, similar to the ``aggressive pre-aggregation'' approach we analyzed and compared against in this
paper.
\lstset{language=SQL, basicstyle=\ttfamily}  
  \begin{lstlisting} [frame=tb,mathescape=true,caption=\texttt{[Q1]} Query for finding the number of customers each supplier \textit{could} reach/ supply parts to per zipcode (TPC-H dataset) ,captionpos=b,label=lst:q1]  %

  SELECT ps_suppkey, c_zipcode, COUNT(*)
  FROM partsupp, lineitem, orders, customer
  WHERE ps_partkey = l_partkey AND
        o_orderkey = l_orderkey AND
        o_custkey = c_custkey
  GROUP BY ps_suppkey, c_zipcode;
\end{lstlisting}

\begin{figure}[t]
 \includegraphics[width=0.28\textwidth]{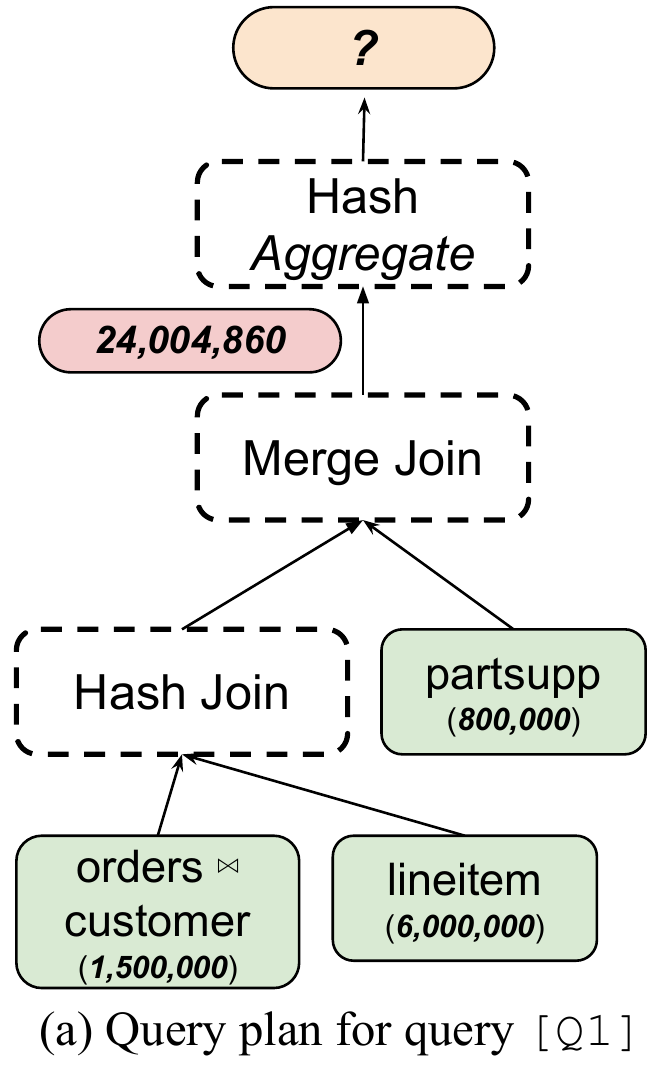}
 \centering
 \caption{Query plan for query \texttt{[Q1]}. Aggregate queries can have very large intermediate results even though the number of output groups could be small}
 \label{fig:motivating-ex}
 \end{figure}

Consider a query like \texttt{[Q1]} in Listing~\ref{lst:q1} over the standard TPC-H dataset of \texttt{parts, customers} and \texttt{suppliers}. The \texttt{lineitem} table includes all orders of parts that were
supplied, the individual parts each order contains, as well as \textit{which supplier} each part was bought from. The goal of \texttt{[Q1]} is to compute the number of
parts that a supplier \textit{could} provide to a certain zipcode, \textit{given} the transaction data that we already have. Note that \texttt{c\_zipcode} isn't a
distinct field in the \texttt{customer} table, but is typically extracted from the \texttt{c\_address} attribute. This type of complex \textit{decision-support} query requires a \textit{non-key} join which could yield very large intermediate results that will be fed as input to the aggregation operator. As shown in Figure~\ref{fig:motivating-ex}, running \texttt{[Q1]} over TPC-H (using scale factor \texttt{SF=1}), the intermediate join result for this query contains over 24 million tuples. The size of the result post-aggregation would be bounded by the number of distinct zip codes times the number of suppliers, and therefore is highly likely to be orders of magnitude smaller than the join result.

Another very common situation where a decision-support query would require a non-key, large-output join is that of \textit{self-joins}, where a table is joined with
itself. For example, on the standard TPC-H dataset, a classic data-mining task is to compute the number of times pairs of parts appeared in the \textit{same order}; this
is a standard frequent pattern mining query. A self-join between the \texttt{lineitem} table and itself (on \texttt{partkey}) is required to compute this.

Another example of queries that require large-output self-joins include \textit{path aggregation queries} in graphs. Any graph stored inside a relational database in the form of a \textit{Nodes} and \textit{Edges} table, is conducive to queries that e.g. count the number of paths that follow a certain pattern in terms of the nodes. If we had an edge table \texttt{Edges(src,dst)}, and a \texttt{Nodes(id,label)}, a query like \texttt{[Q2]} in Listing~\ref{lst:q2}, counts the paths between nodes with certain labels. Queries like this one end up outputting a huge number of intermediate results which constitute the sub-paths for each intermediate stage of the graph traversal.

\lstset{language=SQL, basicstyle=\ttfamily}  
  \begin{lstlisting} [frame=tb,mathescape=true,caption=\texttt{[Q2]} Generic graph pattern counting query, captionpos=b,label=lst:q2]  %

  SELECT n1.label, n2.label, COUNT(*)
  FROM Nodes n1, Edges e1,
       Edges e2, Nodes n2
  WHERE n1.id = e1.src AND
        e1.dst = e2.src AND
        n2.id = e2.dst
  GROUP BY n1.label, n2.label;
\end{lstlisting}


Our main contributions in this paper are twofold; \textit{first}, we propose a new \textit{multi-way} database operator called \joinAgg, which enables the efficient
computation of aggregation queries, \textit{without materializing} any intermediate join results, by computing the \textit{join} and \textit{aggregation} simultaneously.
We describe a novel \textit{general framework} for executing aggregation over conjunctive queries of arbitrary numbers of relations, and numbers of group by attributes that
may be derived from any participating relation, by leveraging a \textit{graph} representation of the underlying \textit{data}. We restrict our formal development to {\bf
    acyclic} queries -- although our algorithms can be adapted to handle cyclic queries, systematically combining our data-level optimizations with the recent work on
    cyclic joins raises complex issues that are beyond the scope of this paper. We implement a prototype of the \joinAgg operator \textit{outside} of the RDBMS and experimentally showcase the benefits of our operator over synthetic and real datasets.

\textit{Second}, we provide a comprehensive complexity analysis of common example queries that benefit from our \joinAgg operator in comparison to executing them using the classical RDBMS model, or other less general techniques such as partial pre-aggregation~\cite{larson2002data} which only looks at reducing intermediate data size at the level of each individual join instead of looking at the join as a whole. We show that in terms of computational complexity \joinAgg is comparable or asymptotically better than those techniques, particularly in the general case of complex acyclic branching join queries. We also show that \joinAgg is overall \textit{better} than those techniques in terms of memory complexity.


The rest of this paper is organized as follows: Section~\ref{sec:preliminaries} introduces some preliminary definitions required later on and provides a high level description of the \joinAgg operator, Section~\ref{sec:datagdef} describes the process with which we load the data into memory in a data graph representation which is stage 1 of the \joinAgg algorithm. Section~\ref{sec:traversal} describes the traversal and result generation stages of the algorithm after defining basic concepts used the traversal process.
Furthermore, Section~\ref{sec:complexity} presents a comprehensive complexity analysis comparing the complexity of \joinAgg with other traditional techniques, Section~\ref{sec:implementation-details} goes deeper into the implementation details of the prototype we implemented, while Section~\ref{sec:experiments} presents the experimental evaluation of the aforementioned prototype.
We finally place this work into context with other related work in Section~\ref{sec:related-work}, and conclude in Section~\ref{sec:conclusion}.

 \lstset{language=SQL, basicstyle=\ttfamily}  
 \begin{lstlisting} [frame=tb,mathescape=true,caption=\texttt{[Q3]} Generic Multi-Attribute Group by Query,captionpos=b,label=lst:q3]  %

   SELECT $A.a$,$B.b$,$C.c$, COUNT(*)
   FROM $R_1~A$, $R_2~J$,$R_3~B$,$R_4~C$
   WHERE $A.j_1$=$J.j_1$ AND $J.j_2$=$B.j_2$ AND $J.j_3$=$C.j_3$
   GROUP BY $A.a$,$B.b$,$C.c$;
  \end{lstlisting}

\section{Preliminaries and Overview}
\label{sec:preliminaries}


  \begin{figure}[t]
  \includegraphics[width=0.35\textwidth]{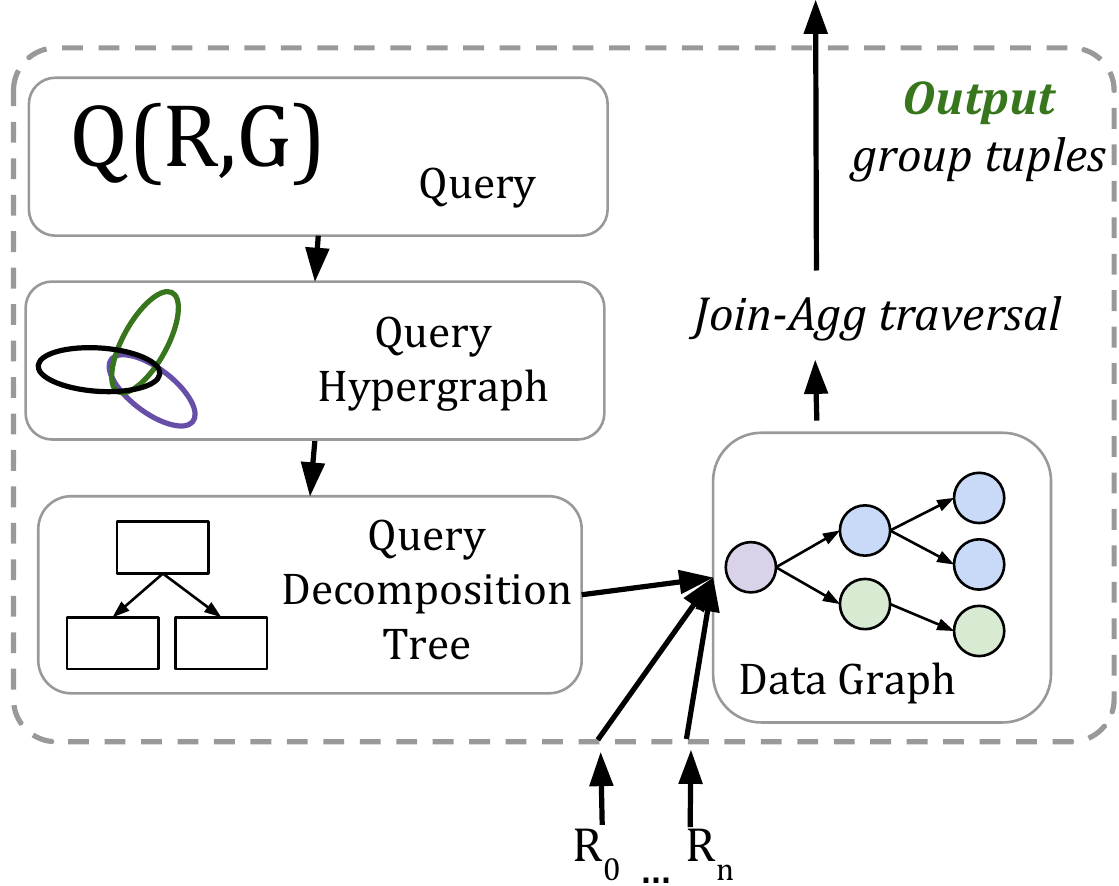}
  \centering
  \caption{Inner-workings of the \joinAgg operator.}
  \label{fig:operator}
  \end{figure}

In this section we formally describe the general framework for efficiently answering queries such as \texttt{[Q3]} in Listing~\ref{lst:q3}. 
Our framework views the join between a series of relations in the form of a {\em graph} structure of interconnected tuples that we call the \textit{data graph}.
For the sake of simplicity, we use the \texttt{COUNT(*)} aggregation function in our explanations and examples. We provide a discussion on how more standard aggregation functions can be supported using the same framework in Section~\ref{sub:other-func}.

\subsection{Preliminaries}
\label{sub:preliminaries}

Let $Q(R,G)$ be an aggregate query over a join between a set of relations $R$, where $G = \{g_1,g_2,...,g_n\}$ is the set of group-by attributes of this query. For now,
assume that we only need to count the number of tuples in each group (\texttt{COUNT(*)}). We assume without loss of generality that a group by attribute  $g_i
\in G$ corresponds to a single attribute in relation $R_i$, and that \textit{none} of the $g_i$ participate in a join condition. We also assume that all joins are natural
joins. All of these can be relaxed easily through standard tuple-level transformations (e.g. if a group-by attribute participates in a join, we can (implicitly) create a copy of that column). As mentioned earlier, we restrict our attention to {\bf acyclic joins} in this paper.

We represent the overall join-aggregation query as a hyper graph $H(X \cup G,E_H)$ where $X$ is the set of all attributes that take part in the \textit{join conditions} between the relations in $R$ and $E_H$ hyperedges, containing one hyperedge $e_{R_i}$
per relation $R_i$, i.e., $e_{R_i} = R_i \cap (X \cup G)$. Note that the only attributes that are relevant here are either join condition attributes, or group attributes--as the result is a set of tuples that represent groups (grouped by $G$).
Let $R_i.x$ denote attribute (or set of attributes) $x$ from relation $R_i$.

For every $e_{R_i} \in E_H$, we partition the attributes of $e_{R_i}$ into two disjoint groups $(x_l, x_r)$. We describe the specifics in Section~\ref{sub:splitting-attr}, but intuitively this is done so as to
reduce the size of the \textit{``data graph''} that we load into memory, while also capturing enough information to execute the query.

\subsection{JOIN-AGG Operator}

We propose a new database operator called \joinAgg that receives a set of input relations $R$ and outputs a single set of result tuples, i.e., after the appropriate grouping and aggregation, as the output.
The decision of whether to use the operator is made by the query optimizer in a cost-based manner; in essence, if at least one of the joins in the query is a
\textit{non-key} join or a join that may result in a large output compared to the input relations, then this new opeartor should be considered. We develop the necessary cost models to make this decision.
When the operator is chosen, instead of conducting a series of binary joins as traditional RDBMSs do, we would instead go through each relation, and load each one into the same in-memory \textit{data graph} which is
then traversed to output the resulting grouped tuples.

Prior to the instantiation phase, the operator creates a {\bf query hypergraph} $H$ that captures the joins in $Q(R,G)$. This query hypergraph is then turned into a {\bf query
    decomposition tree}, which is traversed in order
to transform each individual relation into a set of \textit{edges} in the data graph. Based on the final decomposition, during the execution phase, the operator constructs the edges that correspond to
each relation as the {\bf data graph}. Finally, this in-memory data graph structure is used (and potentially re-used) to directly compute and output the grouped tuples.

\section{Data Graph Representation and Construction}
\label{sec:datagdef}
In this section, we begin with describing how a {\bf query decomposition tree} is constructed and how it is used to split the attributes of each relation into two groups that form the edges of the data graph. We then describe the basic
representation of a {\bf data graph} and explain how it is constructed by loading in relations from the underlying database.

\begin{figure*}[t!]
\includegraphics[width=\textwidth]{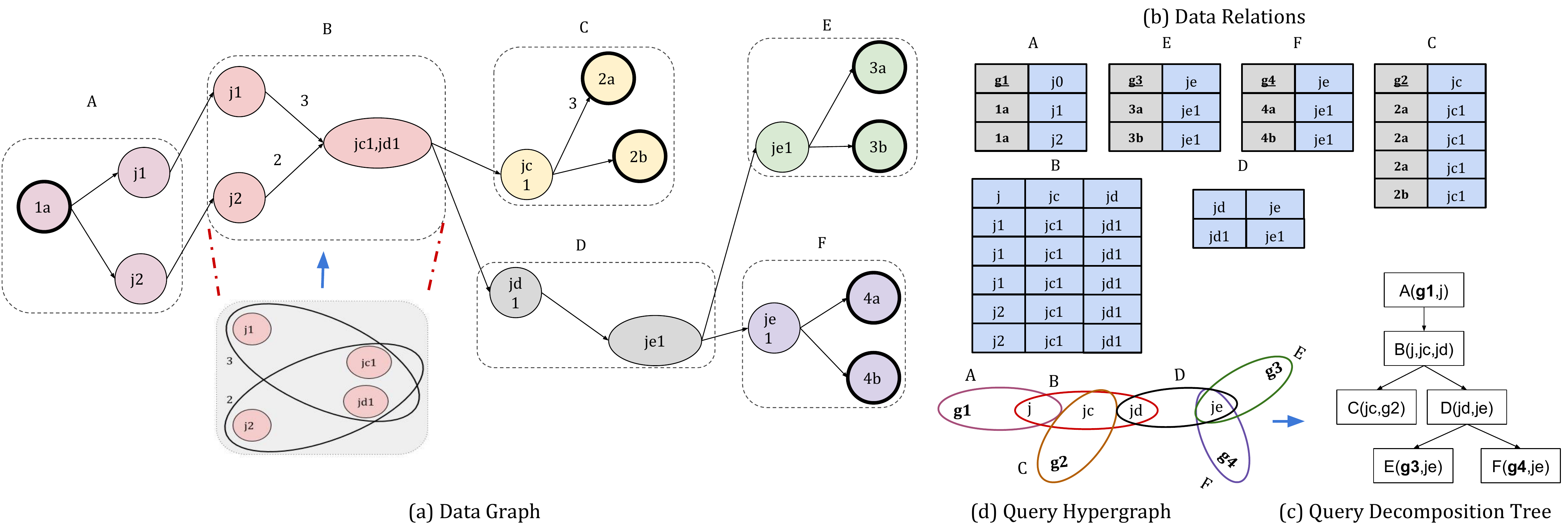}
\caption{A Data graph created by a set of joining relations (after projections have been applied). Relation \textit{B} has multiple attributes as part of $x_r$, which merge into the multi-node \texttt{(jc1,jd1)}. In the relations involved in the join, we have four different group attributes $g_i$, one of which is a source attribute ($g_1$). Node \texttt{1a} is a \textit{source} node,  \texttt{2a, 2b, 3a, 3b, 4a, 4b} are all group nodes, and \texttt{(jc1,jd1)} and \texttt{je1} are both \textit{branching} nodes. The rest are all \textit{intermediate} nodes.}
\label{fig:running-ex}
\end{figure*}

\subsection{Query Decomposition}
\label{sec:decomposition}


A query decomposition of a hypergraph $H$ is defined as a \textit{tree} where each node corresponds to a hyperedge $e_H \in H$. We create a \textit{pure} query decomposition of $H$ where each node in the decomposition directly corresponds to a single relation.
In this work, we focus on acyclic queries, i.e., queries for which there exists a \textit{tree} decomposition~\cite{gottlob2016hypertree}; in future work, we are planning to extend our approach to handle
cyclic queries by combining it with recently proposed techniques for optimal worst-case join algorithms.
%
%
%
%

We construct the query decomposition tree using the standard {\em elimination} algorithm~\cite{tarjan1984simple}. First, we note that, all of the relations that contain \textit{at least one} attribute not present in any other relation must contain a group attribute; we'll call these \textbf{group relations}. We start with one of those as the root of the tree, and traverse the hypergraph in a breadth-first manner to construct a query decomposition tree.
An example of such a decomposition can be seen in Figure~\ref{fig:running-ex}d. Given $H$, to build the query decomposition tree, we can start from any group relation; here we picked $A$. The hypergraph
is then \textit{traversed} in a breadth-first manner starting from $A$ using a standard queue. We start by creating a root node in the decomposition tree for $A$ -- let that be the \textit{current} node.
Then, for every neighbor relation of $A$, if it has not been visited, we add it to the queue. We then pop the queue and add the popped relation as a \textit{child} to the current node in the decomposition
tree. Thereafter we continue with neighbor $B$ which becomes the current node, it is added as a child node to $A$ and so on until the queue is empty.


\begin{figure}[t]
\includegraphics[width=0.35\textwidth]{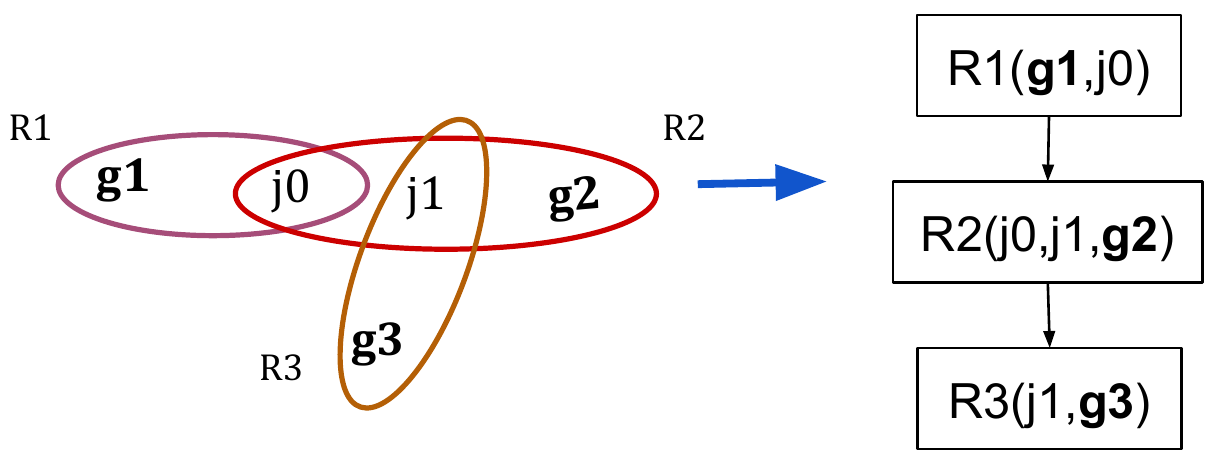}
\centering
\caption{Derivation of a Query Decomposition tree from a Query Hypergraph}
\label{fig:small-decomposition}
\end{figure}

\subsection{Splitting Attributes}
\label{sub:splitting-attr}

As previously mentioned, the attributes of each relation are partitioned into a pair of attribute sets $(x_l,x_r)$. This is done in order to properly view every relation as a set of \textit{edges} for the data graph--an edge has two entities it connects, here it connects $x_l$ and $x_r$. To do this, we simply traverse the query decomposition tree starting from the root. As
 we discuss in further detail later on, this splitting mechanism is a \textit{data reduction} mechanism (similar to {\em pre-aggregation}~\cite{larson2002data}) for reducing the input data as much as possible before the query is executed.

To construct these pairs, we traverse the decomposition tree in order to partition the set of attributes for each relation into a $(x_l,x_r)$ pair.
In a slight abuse of notation, let $R_i$ also represent the \textit{set of attributes} relevant to the query from relation $R_i$. We start the traversal at the root; let that be $R_S(g_0,a_1,...,a_n)$,
   for which we set $R_S.x_l=\{g_0\}$. Afterwards, we set $R_S.x_r= \bigcap{C_S}$ where $C_S = R_S.children() \cup \{R_S\}$ is the \textit{set of attributes} in the children relations of $R_S$ in the decomposition tree, plus (union) $R_S$ itself. Next, for every relation $R_i \in R_S.children()$, if $R_i$ \textit{is not} a group relation, we set $R_i.x_l = parent(R_i).x_r$, and again $R_i.x_r= \bigcap{C_i}$.
If $R_i$ \textit{is} a group relation, we set $R_i.x_l = R_i \setminus \{g_i\}$ and $R_i.x_r = \{g_i\}$. As we will later describe, nodes created from group attributes need to be \textit{sinks} in the DAG structure that we will be building.
This same process is recursively executed on all relations in $R_i.children()$.

Below we provide a few standard examples of how the aforementioned algorithm is used to split the attribute sets.

\begin{exmp}
  Looking at the decomposition tree in Figure~\ref{fig:running-ex}c,  a simple and common case is that of \texttt{D}. Because \texttt{D}'s parent relation is \texttt{B}, we have $x_l = B \cap D = \{jd\}$. Also, the children of \texttt{D} are \texttt{E,F}, therefore $x_r = D \cap E \cap F = \{je\}$. In the data graph constructed later, a single node will be created for each value in $\pi_{jc}(D)$ and for each $\pi_{je}(D)$.
\end{exmp}

\begin{exmp}
  Again regarding the tree in Figure~\ref{fig:running-ex}c, for relation \texttt{B}, its $x_l$ value will be $B \cap A = \{j\}$, and $x_r = B \cap C \cap D = \{jc,jd\}$. This multi-valued $x_r$ value means that a \textit{multi-node} will be created for each value in $\pi_{jc,jd}(D)$.
\end{exmp}

\begin{exmp}
  Looking at the decomposition tree in Figure~\ref{fig:small-decomposition}, for relation $R2$, since it's a \textit{group} relation (one that contains at least one group attribute), we \textit{always} split its attributes by setting the group attribute as $x_r = \{g2\}$ ,and $x_l = R2 \setminus \{g2\} = \{j0,j1\}$.
\end{exmp}

\subsection{Data Graph Representation}
\label{sec:data-graph-def}
Next, we formally define the data graph representation for a join-aggregate query. For a given query $Q$, a \textit{data graph} $G_Q(V, E)$ captures the underlying data in the relations and the interconnections between those data elements.
Let $n \in V$, and $e = (n_l,n_r) \in E$ the nodes and (directed) edges respectively that make up graph $G_Q$.

\textbf{\ul{Relation \& Node Types:}} At a high level, we partition the relations $R$ in four (overlapping) groups:  $R_S, R_B, R_J, R_G$, which dictates how the corresponding nodes are handled during execution of the \joinAgg algorith:

\begin{enumerate}
  \item $R_G= \{G_0,G_1,...G_n\}$ denotes the \textit{group} relations in $R$, each containing a group by attribute $g_i$.
  \item $R_S$ is the \textit{source} relation $R_S \in R_G$, which we choose to start the computation from. The algorithm we develop is based on a \textit{traversal} of the data graph. As we describe in
  Section~\ref{sec:traversal}, nodes originating from the source relation are the \textbf{anchors of the traversal} and therefore do not get visited multiple times. As a result, no additional data needs to be maintained for them.
  \item $R_B =\{B_0,B_1,...,B_n\}$ a set of \textit{branching} relations in $R$. A relation is marked as a branching relation if its corresponding node in the \textit{query decomposition tree}: (a) has \textit{more than one} child and therefore branches out the join execution, or (b) is \textit{not} a leaf node or the root node, \textbf{and} \textit{is} also a group relation. In other words, either the tuples in these relations need to sequentially be joined with multiple tables, one at a time in the context of the overall conjunctive query, or there is a grouping attribute in the relation that needs to be separated out so that we can exploit {\em caching}
  effects (as discussed later). An example of a type (a) branching relation is $D$ in Figure~\ref{fig:running-ex}d, and a type (b) would be relation $R2$ in Figure~\ref{fig:small-decomposition}.
  \item $R_J = \{J_0,J_1,...,J_n\}$ a set of intermediate relations in $R$. These are relations that only have exactly \textit{one child and one parent} in the query decomposition tree, and are \textit{not} group relations.
\end{enumerate}

Consequently, there are four types of nodes in the data graph, each originating from its adjacent relation type: source nodes $(n_s)$, group nodes $(n_g)$, intermediate nodes $(n_j)$, and branching nodes $(n_b)$. Each of the aforementioned relations all portray a pair of attributes $(x_l, x_r)$ that are relevant to the query. As discussed in Section~\ref{sub:step-1}, source nodes are
always loaded from the $x_l$ attribute in the \textit{source} relation, while group and branching nodes from the $x_r$ attribute in their respective relations. Group nodes are always \textit{sinks} in the data graph, while we made the choice to always set branching nodes on the right hand side of the split for the sake of simplicity. All remaining nodes in the data graph are intermediate nodes.  

Note that a relation can have \textit{multiple types} (e.g., $R_S$ is both a source relation and a group relation). Similarly, based on the specific query, a branching relation can also be a group relation if it satisfies the criteria for both relation types simultaneously. This would occur for instance if a relation includes a group attribute $g_i$, but \textit{also} joins with more than two relations in the query described by a hypergraph $H$. The \textit{nodes} derived from relations with multiple types naturally also inherit the same set of node types.

Attribute \textit{splitting} (Section~\ref{sub:splitting-attr}) enables us to conduct a \textit{pre-aggragation} step in which we group tuples with the same values (after projection) into a single edge with a \textit{multiplicity} value. For example, in regards to relation $C$ in Figure~\ref{fig:running-ex}, splitting attributes this way not only allows us to pre-aggregate $(2a, jc1)$, but to also only load in a single node for $jc1$.

\subsection{Mapping Relations to a Data-Graph}
We now formally describe how we map rows in the underlying relations to nodes and edges in the data graph.
Let $\pi^*_{x_1,x_2,..,x_n}(R_i)$ denote the set of values of attributes $x_1,x_2,..,x_n$ in relation $R_i$, where $\pi^*$ indicates bag semantics for the projection.
Also, let $\pi_{x_j}(R_i)$  denote the set of unique values of the attribute $x_j$ in relation $R_i$, and let $X_i$ denote the set of attributes in relation $R_i$ that take part in a join condition.
We create the nodes in the data graph in two simple steps; for every relation $R_i \in R$:
\begin{enumerate}
  \item We create a \textit{hyperedge} for every $(x_l,x_r)$ tuple in relation $R_i$, as is seen in Figure~\ref{fig:running-ex} (hyperedges with only $2$ nodes are shown as regular directed edges). Every hyperedge describes a set of values from attribute sets $x_l \cup x_r$. A unique value in attribute set $x_i$ corresponds to a single node in such a hyperedge.
  \item For every set of nodes in the \textit{intersection} of a set of hyperedges, create a \textit{multi-node} that includes \textit{all} values in the intersection (also shown in Figure~\ref{fig:running-ex} for relation $B$). The result is a set of regular directed edges between nodes and/or multi-nodes. For all purposes moving forward, multi-nodes function exactly the same way as regular nodes in the data graph. In general, the node created from an attribute set $x_i$ is simply $n_i$, and its set of values are denoted as $v_i$.
\end{enumerate}

We now define the \textit{edges} in $G_Q$. Let $m_e$ denote the \textit{multiplicity} of an edge $e \in E$. The multiplicity of an edge is a numeric value associated with each edge and is defined as the number of times the tuple that edge corresponds to exists in the relation. $G_Q$ contains a directed edge $(n_l,n_r) \in E$ iff one of the following applies:

\begin{enumerate}
  \item There exists a tuple $(v_l, v_r) \in \pi_{x_l,x_r}(R_i)$.
    If $A = \{ (a,b) \in \pi^*_{x_l,x_r}(R_i) : a = v_l \wedge b = v_r \}$, the set of tuples in $R_i$ with values $(v_l, v_r)$ in $R_i$, the multiplicity $m_{(n_l,n_r)} = |A|$.
    \item A tuple in $R_i$ joins with one in $R_j$ on attribute $x_{join} = X_i \cap X_j$, such that $v_{join} = v_r = v_l$. In this case, the multiplicity of the edge is always $m_{(n_l,n_r)} = 1$.
    An example of such an edge is $(n_{A.j1}, n_{B.j1})$ in Figure~\ref{fig:running-ex}a.
\end{enumerate}

For the sake of simplicity, and without loss of generality we can assume that any $x_i$, corresponds to a single attribute, i.e. relations only join with one another through single attribute join conditions.
In practice, $x_i$ can be a set of attributes, in which case $v_i$ (the value for node $n_i$) would constitute of a bag of values and be described as a multi-node in the data graph. Formally, in that case we simply have that $(n_l,n_r) \in E :$ $x_l \cap x_r \neq \emptyset$ where $x_l = X_i$ and $x_r = X_j$ and $v_l \cap v_r \neq \emptyset$.


\subsection{Join-Agg Stage 1: Loading Data Graph}
\label{sub:step-1}
To summarize, the input to the overall load process is the hypergraph $H$. We initially need to partition all sets of relation attributes to $(x_l,x_r)$ pairs, by first creating a \textit{query
    decomposition} of $H$, and then using that decomposition to do the partitioning.

The data graph is then loaded into memory by simply scanning the input relations, sorting them, and creating nodes as described above. If there are any attributes in the input relations that don't
participate in the query, we push down appropriate projections (without duplicate elimination) to the underlying database to minimize the amount of data transferred over the network.

\section{Traversing The Data Graph}
\label{sec:traversal}
In this section, we describe our algorithm that computes the aggregated groups of such a query $Q$, by traversing a data graph $G_Q$. 
For the sake of simplicity we will focus on the query that \textit{counts} the number of tuples in each group and discuss how it is generalized in Section~\ref{sub:other-func}.

The high level idea behind \joinAgg is it to traverse the data graph, which represents the underlying data being joined, starting for one source node at a time and maintain certain partial aggregate values (in this case, counts) \textit{at all reachable group nodes} in each iteration.
We can later \textit{combine} these values in order to obtain the final aggregate value of each group, instead of materializing the join at any point. The way this happens at a high level is by propagating the counts along the data graph, starting from each unique source node, to the group nodes, while keeping track of certain path information (which we refer to as \textit{path-id}s) along the way. These path-ids allow us to figure out which counts are derived from which paths in the data-graph and enable us to properly combine them to compute the correct count for each group.

\subsection{Definitions \& Axioms}
\label{sub:definitions}

Before we formally describe our general algorithm for executing these queries over a data graph $G$, we enumerate a few core definitions and axioms for concepts that we be regularly reference in the algorithm description. The execution algorithm we propose revolves around traversing the data graph and maintaining certain information along the way in order to \textit{directly} output the groups in the result.

\begin{definition}
  \label{def:rtree}
  A \textit{rooted tree} (also formally known in the context of directed graphs as an \textit{arborescence}), is defined as a directed subgraph that consists of a tree, with a single root node, therefore containing \textit{exactly one} path between that root node and every leaf node.
\end{definition}

\begin{definition}
  \label{def:count}
 Let $C(n_1,n_2)$ denote a \textit{count} between nodes $n_1$, $n_2$. We conceptualize the traversal of the data graph as equivalent to \textit{conducting joins} between the tuple that each element of the data graph represents, thus generating new tuples which we want to avoid materializing. A count represents the \textit{number of tuples generated along all paths} between node $n_1$ to $n_2$. Any such path \textit{cannot} include a branching node ($n_1,n_2$ may themselves be branching nodes, but there cannot exist a branching node in any path between them).
 More formally, using Axiom~\ref{ax:path} we have $C(n_1,n_2) =$ $\sum_{i=0}^{k} (|n_1\rightarrow N_{j_i} \rightarrow n_2|)$, where $N_{j_i}$ the set of intermediate nodes in one of the $k$ unique paths between $n_1$ and $n_2$.
\end{definition}

\begin{definition}
  \label{def:pathid}
  Let $p=[v_{b_0},v_{b_1},..,v_{b_n}]$ denote a \textit{path-id}. A path-id is a \textit{list of branching nodes} found in a unique path from a source node $s$ to a group node $g_i$. We maintain path-ids in order to logically re-construct all possible rooted trees which have $s$ as their root, and include all group nodes $n_{g_1},n_{g_2},..,n_{g_i}$ in their leaves in order to compute the number of tuples within each group $(v_s, v_{g_1}, v_{g_2}, ..,v_{g_i})$. Path-ids are unique identifiers for unique paths in the data graph, and are always paired with a path-id \textit{count} described below.
\end{definition}

\begin{definition}
  \label{def:pathid-count}
   A \textit{path-id count} denoted by $C_{p_i}$, is defined as the count between two branching nodes $n_{b_i},n_{b_j}$ and is  equal to $C(n_{b_i},n_b{_j})$, where path-id $p_i
   =[v_{b_0},...v_{b_i},v_{b_j}]$. The path-id maintains information about the rooted trees this certain path is part of. The path-id count itself however represents the count between the \textit{last
       two} branching nodes in the path-id (even though the path-id might include more than two branching nodes). In the case where $|p_i|=1$ then $C_{p_i} = C(s,n_{b_j})$ where $s$ a source node. The intuition here, thinking about this from a query processing perspective, is that we need to keep track of \textit{how many tuples were generated} at the point where a relation joins with more than one other relation. Once we join with one of the relations, we need to go back and join with the rest one at a time. In order to do that properly (without actually materializing the join result) we need to know how many tuples were generated at that time in the query before it branches off to multiple joins.
\end{definition}

\begin{definition}
  \label{def:group-count}
  A \textit{group-node count} denoted $c_i =C(n_{b}, n_{g})$ is the count between the last branching node $n_{b}$ of a path, and a group node $n_{g}$.
  Intuitively, a group-node count $c$ represents the number of tuples generated by joining the tuples in the underlying relation that contain value $v_b$, with all intermediate tuples, and then also joining them with all tuples that contain $v_g$.
\end{definition}

\begin{definition}
  \label{def:c-pair}
   A \textit{c-pair} denoted by $P = (p,c)$, is a pair consisting of a path-id and a group-node count. These pairs are recorded at every group node during the traversal of the data graph described in the algorithm in Section~\ref{sub:step-2}.
\end{definition}

\begin{axiom}
  \label{ax:path}
  Let $|n_1 \rightarrow N_j \rightarrow n_2|$, denote the number of tuples generated when there exists a path from $n_1$ to $n_2$ which includes a set of in-between (either branching or intermediate) nodes $N_j=\{ n_{j_1}, n_{j_2},...,n_{j_n}\}$.
  By definition of the data graph (Section~\ref{sec:data-graph-def}), there \textbf{must} exist tuples $t_1 = (v_1,v_{j_1}), t_2 = (v_{j_1},v_{j_2}),..., t_n=(v_{j_n},v_2)$, (where a tuple $t_i$ appears  $m_{(n_l,n_r)}$ times in its corresponding relation),  such that $\{(n_1,n_{j_1}),(n_{j_1},n_{j_2}),..,(n_{j_n},n_2)\} \in E$.
  The number of tuples generated is $|t_1 \Join t_2 ...\Join t_n| =m_{(n_1,n_{j_1})} * m_{(n_{j_1},n_{j_2})} *..* m_{(n_{j_n},n_2)} = |n_1 \rightarrow N_j \rightarrow n_2|$, and is derived by taking the product of all edge multiplicities along the path.
\end{axiom}

\subsection{Join-Agg Stage 2: Traversal and Multiplicities}
\label{sub:step-2}
Stage two of this algorithm traverses $G_Q$ in a depth first fashion starting from each \textit{source node}, properly keeping track of the cumulative edge  multiplicity along the way, and finally setting the appropriate \textit{c-pairs} at all reachable group nodes.

A depth first traversal starting from source node $n_s$ to the rest of the group nodes consists of multiple different rooted trees, each of which ends up at a potentially different set of leaf nodes (group nodes). Every rooted tree that reaches \textit{exactly one of each type} of group node, corresponds to a tuple (or set of identical tuples) in the result of the join $\calR$. The purpose of this algorithm is to \textit{count} the number of such rooted trees that each combination of group nodes has \textit{in common}.

The result of the traversal step is a set of lists $L$, containing one list of \textit{c-pairs} associated with each group node $n_{g}$ reachable from $n_s$ -- let $l_{n_g}$ denote each such list. Again, $(p,c)$ denotes a \textit{c-pair}, comprising of a path-id $p$, and a group-node count $c$. There is also a path-id count $C_{p}$ associated with each unique path-id (we define the terms path-id, path-id count and group-node count in Section~\ref{sub:definitions}).

We now outline the process that traverses the data graph and sets the appropriate c-pair lists at every group node $n_{g_i}$.
We start at a source node $n_s$, and conduct a DFS traversal. Let $p_c$ denote the current path-id, $c_c$ denote the current count, and $n_c$ the current node being visited. Also let $n_{c_i}$ denote the $i$'th neighbor of $n_c$, and $m_{(n_c,n_{c_i})}$ denote the multiplicity of the edge between them.

We now define a recursive \textit{visit($n_c$)} function: if $n_c$ is a group node, record $(p_c,c_c) \rightarrow l_{n_c}$ , and return.
This is the base case of the recursion. If $n_c$ is not a \textbf{group} node, for each $n_{c_i} \in out(n_c)$, if $n_{c_i}$ is a \textbf{branching} node, update $c_c = c_c * m_{(n_c,n_{c_i})}$ with the current neighbor's multiplicity, append $n_{c_i}$ to the path-id $p_c$, and reset $c_c = 1$. The reason we reset the current count is because we now need to keep track of the count along the \textit{new path} since we encountered a \textbf{branching} node.
If $n_{c_i}$ has already been visited by this traversal (the traversal starting from $n_s$), simply update that path-id count to $C_{p_c} = C_{p_c}+c_c$ and return. Next, recursively \textit{visit} every $n_{c_i} \in out(n_c)$. This can be seen as a form of computation \textit{caching}. If we've been through a path in a current traversal, we don't need to go through it again, whereas in traditional execution, this path would be computed multiple times (in the form of joining tuples).


\subsection{Join-Agg Stage 3: Result Generation}
\label{sub:step-3}
Finally, we end up with a list $l_{n_g}$ for every $n_g$ reachable from $n_s$ -- let $N_{n_s} = [n_{g_0},n_{g_1},...,n_{g_k}]$ denote this set of group nodes. We utilize these lists in order to generate the final result groups. The \textit{intuition} behind this process is that a group $(v_s, v_{g_0},v_{g_1},...,v_{g_k})$ in the final result will only have a non-zero count value iff there is \textit{at least one} rooted-tree in the data graph with $n_s$ as the root, and $N_{n_s}$ as leaves. Every c-pair set during the \textit{traversal} stage of the algorithm will contain a path-id that is part of such a rooted-tree. There is a count computed for every such rooted-tree. The goal of this stage of the algorithm is to use these c-pairs set at every $n_g$ in order to re-construct all of the rooted-trees that contribute to the result, and finally compute the \textit{sum} of all of their counts. That sum is equal to the size of the output group.

First, we separate the group nodes reached by the traversal into a set of \textit{buckets}. The \textit{combination} of all c-pairs found in all nodes in $N_{n_s}$, will result in the final count for the group $(v_s,v_{g_0},v_{g_1},...,v_{g_k})$; if this count is non-zero, the group is output to the final result. We will now properly explain how this \textit{combination} of c-pairs is conducted.

We partition $n_{g} \in N_{n_s}$ nodes into $|R_G|$ \textit{buckets}, one for each group relation $G_i \in R_G$, by adding a node into a bucket $B_i$ if it was derived from group relation $G_i$. Let $B$ denote this set of group node buckets.
Next, for each bucket $B_i \in B$, we output a list of tuples $F_i$ that we will combine in order to generate the final result. The way we output these tuples is the following: For each node $n_i \in B_i$, for every c-pair $P_i=(p_i,c_i) \in l_{n_i}$, we output a tuple $(n_i,(p_i,c_i)) \rightarrow F_i$, so that we keep track of which group node each c-pair in $F_i$ came from.
Let $F$ denote the set of lists output from all buckets in this step.

Lastly, in order to construct and aggregate all distinct groups that are in the final output and their associated counts, we conduct a \textit{prefix-join} (denoted as $\Join_{\sim}$) of the lists $F_i \in F$ on the \textit{path-id} in a pair-wise fashion. In this prefix-join, two tuples match if their path-ids  \textit{share a common prefix}.

More specifically, let $\sim$ define a binary relationship between path-ids, that indicates they share a common path prefix. Let $p_1,p_2$ path-ids where $l_1,l_2$ are their respective lengths, and $l_1 \leq l_2$.  We say that $p_1 \sim p_2$ iff $p_1[0..l_1] = p_2[0..l_1]$.

Therefore, for every tuple in $F_0 \Join_{\sim} F_1 \Join_{\sim} ... \Join_{\sim} F_i$, we compute a value that will \textit{contribute} to a group in the final result.
Say we're computing $F_1 \Join_{\sim} F_2$; Let a tuple $f_1=(n_1,(p_1,c_1)) \in F_1$ and $f_2=(n_2,(p_2,c_2)) \in F_2$.
If $p_1 \sim p_2$ we output $f_3=(\{n_1,n_2\},(p_i, c_3))$ where $c_3 = C_{p_1} * C_{p_2} * c_1 * c_2)$ and  $p_i = p_1$ iff $|p_1|<|p_2|$ or $p_i = p_2 $ iff $ |p_1|>|p_2|$ i.e. the path-id with the smallest length. We only multiply the result with the path-id count of each unique path-id, \textit{once} --
if e.g. we joined $f_3=(n_3,(p_i,c_3)) \Join_{\sim} f_4=(n_4,(p_i,c_4))$ (a tuple with the exact same path-id), the output tuple would be $f_5=(\{n_3,n_4\},(p_i,c_5))$ where $c_5 = c_3 * c_4 * C_{p_i}$.

For every iteration of the algorithm, we start from a source node $n_s$, we end up getting an $F$ set of c-pairs, out of which we output all $f_i$ tuples resulting from the prefix-join described above iff they have non-zero counts. After the end of step 3, we will have output all groups, that have any combination of values where every value comes from a different group relation.

\begin{figure}[t]
\includegraphics[width=0.40\textwidth]{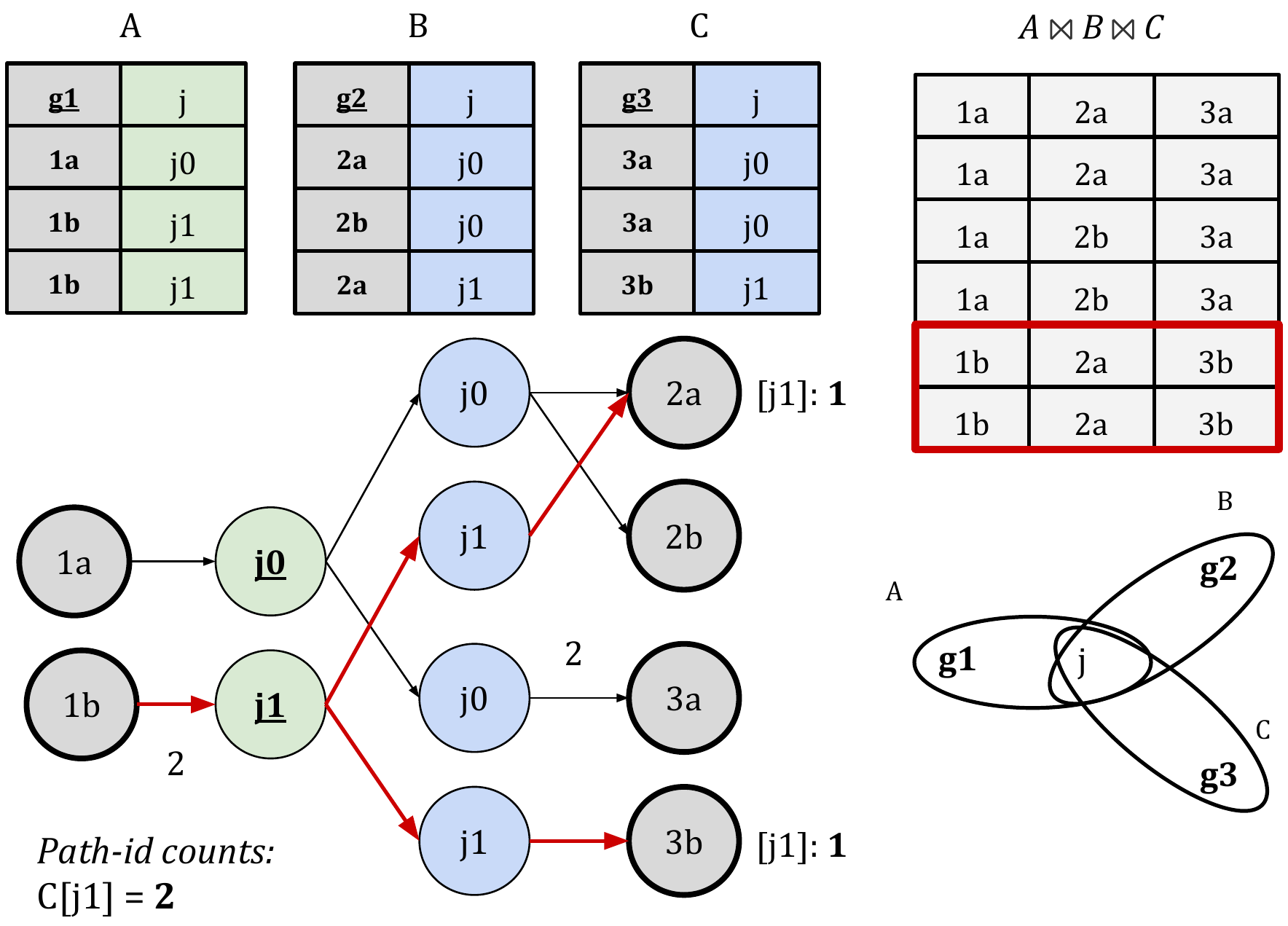}
\caption{A rooted tree in the data graph corresponds to at least one tuple in $\calR$ that contain the values at the root and the leaves of the rooted tree (the source node and the group node values)}
\label{fig:simple-ex}
\end{figure}

\begin{exmp}
  Consider the example on Figure~\ref{fig:simple-ex}, showing the data graph and the join result $\calR$ for this query. The red arrows showcase an example of a \textit{rooted tree}, with source node \texttt{1a} as its root. Every possible rooted tree in $G_Q$ which includes one of each type of group node in its leaves directly corresponds to a tuple in the join result $\calR$. Since the idea is to \textit{avoid} materializing $\calR$, we instead traverse this graph, and set a c-pair at every group node every time it is visited, identifying the path that reached a given node by its unique \textit{path-id}.
  Here we can see that for group node $2a$, its c-pair list is $l_{2a} = \{([j1], 1)\}$, and for $3b$ we have $l_{3b} = \{([j1], 1)\}$ accordingly. We will transform these lists into sets of tuples $\{F_1, F_2\}$ where $F_1 = \{(2a, ([j1], 1))\}$ and $F_2 = \{(3b, ([j1], 1))\}$.
  We compute the prefix-join $F_1 \Join_{\sim} F_2$, which outputs the tuple $f_3$ with the value $(2a, 3b)$ and the count $1 * 1 * 2$ (the count of $f_1$ times the count of $f_2$ times the path-id count for the path-id $[j1]$). Finally, $(1b,2a,3b), 2$ is output. There is such a tuple computed \textit{for every rooted-tree in the data graph} that has $1b$ as its root, and $2a,3b$ as its leaves. The sum of the counts for every unique group is equal to the size of the group in the final result.

\end{exmp}

\subsection{Other Aggregation Functions}
\label{sub:other-func}
The list of basic aggregation functions supported by most SQL execution engines includes \texttt{COUNT,SUM, MIN,MAX} and \texttt{AVG}. We argue that our ability to execute \texttt{COUNT} without outputting individual intermediate results generalizes directly to the rest of these basic aggregation functions.

\topic{\texttt{SUM}} Firstly, \texttt{COUNT} can be thought of as a special case of \texttt{SUM}, if we assume that every single tuple in the group includes an attribute for which the value is always 1. If the value of such an attribute is not 1, while executing the query , we can simply keep track of the running sum of tuples, that include the attribute values being summed over, instead of just the running multiplicity of generated tuples. The sum would then need to be multiplied by the count for a specific group, which would then output the correct result.

\topic{\texttt{MIN, MAX}} These two functions would only require keeping track of a single value and do not require maintaining counts at all.

\topic{\texttt{AVG}} This requires keeping track of the \textit{sum} of the certain combination of attribute values that need to be averaged over, as well as the count that the current version of our algorithm is maintaining.

\section{Complexity Analysis}
\label{sec:complexity}

Here, we provide a high-level analysis of the computational complexity of executing a join-aggregate query, with the goal of showing the asymptotic benefits of our approach. We make several simplifying assumptions for clarity. For any relation $R_i$ that we use in the examples below, we make a \textit{uniformity} assumption about all join condition attributes. Moreover, any join between relations in the below examples are natural joins. Again, let $\pi_{g}(R_i)$ denote the domain of values for attribute $g$ in relation $R_i$.
We assume that all relations $R_i$ in any example are of a constant size $n$.

We contrast the time and space complexity of the algorithm \textit{traditional RDBMSs} would use to compute join-aggregate queries, our \joinAgg operator, and an idealized pre-aggregation approach. We do this by choosing three simplified example queries. For the traditional RDBMS execution we assume that a \textit{sort-merge join} operator is used to compute the join between any two joining relations and that \textit{neither} of the joining relations have indexes on the join condition attribute. In most modern database systems, a \textit{hash-join} operator of some sort would be chosen by the query optimizer, but only if the optimizer accurately estimates the amount of memory required for storing the hash-table given the amount of memory available. In terms of how \textit{aggregation} is performed in the RDBMS, we assume again that a \textit{sort-aggregate} operator is used instead of a \textit{hash-aggregate} operator.

It's also important to note that we are loading in a node in the data graph for every unique value in each distinct relation. For the sake of simplicity we assume \textit{set} semantics for any
relation, i.e., we assume that every tuple in a relation is unique. 
Note that there are various optimizations possible for the first two examples below, which nonetheless don't change the asymptotic complexity of the algorithm.

\begin{figure}[t]
\includegraphics[width=.5\textwidth]{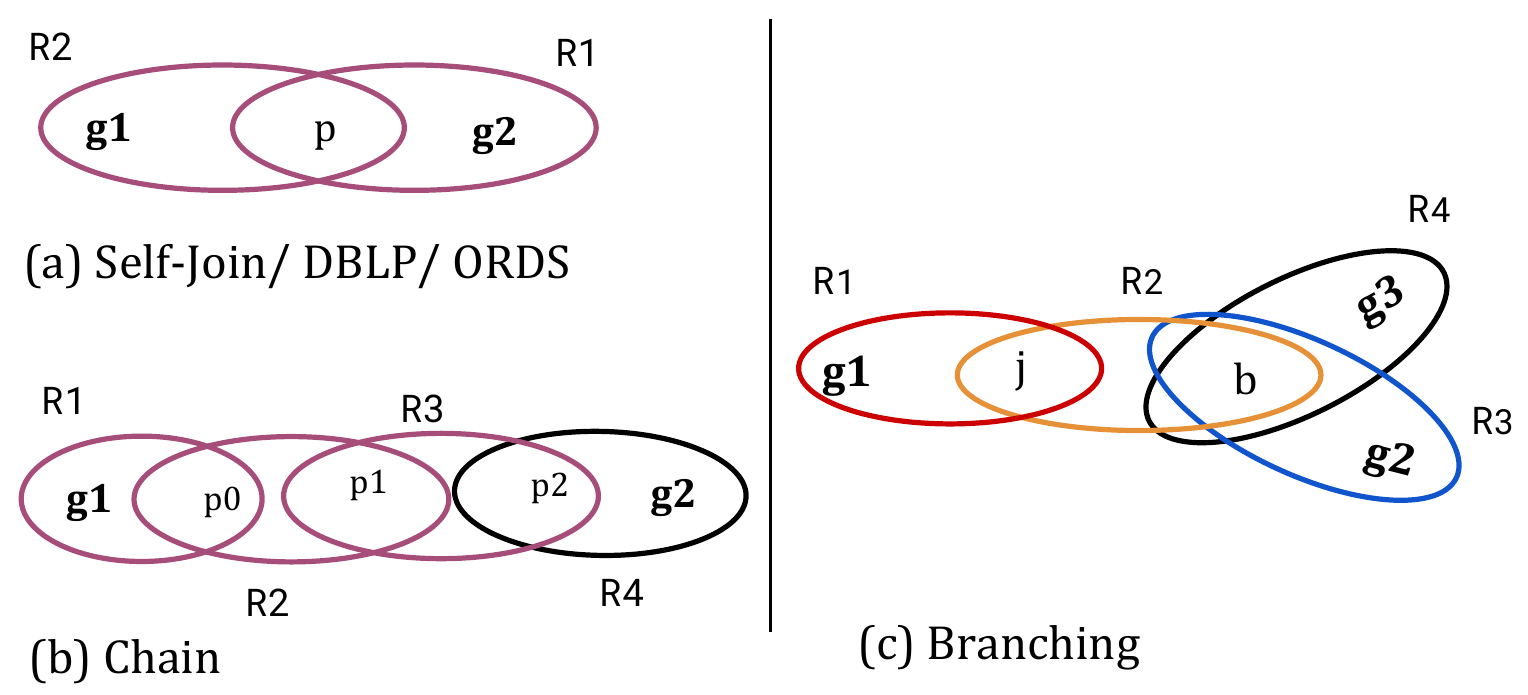}
\centering
\caption{Hypergraphs of example queries}
\label{fig:example-hypergraphs}
\end{figure}

\topicul{Self-Join}
Here, for a single relation $R(g,_,p)$, the join-aggregate query is to do a self-join on $p$, and a group-by on the two copies of the attribute $g$ (shown in Figure~\ref{fig:example-hypergraphs}a). More specifically, let the two copies of $R = \{
R_1(g_1,_,p), R_2(g_2,_,p) \}$, and let $G = \{ g_1, g_2 \}$, giving us the \joinAgg query $Q(\{R_1, R_2\},G)$.

\ul{Traditional RDBMS:} A traditional RDBMS would compute the join and then aggregate the results. Let $\pi_{g_1}(R_1) = \pi_{g_2}(R_2) = a$ and $\pi_p(R_1) = \pi_p(R_2) = b$.
The \textbf{join} computation requires us to first sort both of the relations ($O(2nlogn)$), then compute the join between them and output all result tuples. The complexity of computing the join will be
equal to the number of output tuples in the result of the join, which is $\frac{n^2}{b}$. Overall, the join process takes $O(2nlogn + \frac{n^2}{b})$ steps. The \textbf{aggregation}
process that follows the join requires a sort of the join result, giving us a total time of $O(\frac{n^2}{b}\log(\frac{n^2}{b}))$.

\ul{Join-Agg Operator:} The \textbf{number of vertices} in the data graph $|V| = 2a + 2b$. The \textbf{number of edges} in the data graph $|E|$ is at most $2ab$.
At the \textbf{traversal stage} of the algorithm, we need to conduct a full \textit{dfs} traversal of the graph for every \textit{source} node, of which we have $a$ here. A single \textit{dfs}
requires $O(|V| + |E|)$, therefore overall, we will have: $O(a*(|V| + |E|)) = O(a*(2a + 2b + 2ab)) = O(a^2b)$. Since there is no merging step for this query, the \textbf{result generation} requires
a pass over the reachable group nodes, and there may be at most $a^2$ different results.
Overall we have a time complexity of $O(a^2 + a^2b) = O(a^2b)$.

\ul{Pre-aggregation:} For this query, we can do a partialy pre-aggregation on $R$ by aggregating on $\{g, p\}$, thus reducing its size to at most $ab$. The total execution time then reduces to
$O(a^2b \log(a^2b))$, and is thus comparable to the \joinAgg operator runtime. However, the maximum memory consumed by this approach at any point is $O(a^2b)$, whereas the \joinAgg operator consumes at most $O(ab)$ memory.

\ul{Comparison:} It is easy to see that, the \joinAgg operator performs better asymptotically than the traditional approach if $ab < n\log(n)$, i.e., if the number of unique values of $g$ and/or the number of unique values of $p$ is
small relative to the relation.

\topicul{Chain Join}
Next we consider a simple chain join between four relations, $Q = R_1(g_1,\_,p_0) \Join R_2(p_0,\_,p_1) \Join R_3(p_1,\_,p_2) \Join R_4(p_2,\_,g_2) $ (shown in Figure~\ref{fig:example-hypergraphs}b), still maintaining $2$ group attributes in total.

\ul{Traditional RDBMS:} Computation of the \textbf{join} result is again dominated by the generation of the result tuples, and requires 
$\frac{n^4}{b^3}$ steps. 
For \textbf{aggregation}, we again would sort the join result and scan it to output the result groups, overall requiring $O(\frac{n^4}{b^3}log(\frac{n^4}{b^3}))$.

\ul{Join-Agg Operator:} The \textbf{number of vertices} in the data graph $|V| = 2a + 6b$. The \textbf{number of edges} in the data graph $|E|$ is at most $4ab$. Similarly as the above case, the
\textbf{traversal stage} will take $O(a*(|V|+|E|)) = O(a* (2a + 6b + 4ab))$. Overall, we again have the total time complexity of $O(a^2b)$. 

\ul{Pre-aggregation:} With aggressive pre-aggregation over the input relations and all intermediate results after they are generated, the time complexity of the join-at-a-time approach can be reduced to
$O(a^2b \log(a^2b) + ab^2 \log(ab^2))$. However, the memory consumption of this approach reaches $O(max(a^2b, ab^2))$ at various points during execution.


\ul{Comparison:} As our experimental results also validate, the benefits of a single operator are clearly apparent here, with potentially very large gains coming from more careful and combined evaluation.

\topicul{Chain Join w/ 4 Grouping Attributes}
Next we consider a chain join between four relations, $Q = R_1(g_1,\_,p_0) \Join R_2(p_0,\_,g_2,p_1) \Join R_3(p_1,\_,g_3,p_2) \Join R_4(p_2,\_,g_4) $, but with a total of 4 grouping attributes.

\ul{Traditional RDBMS:} Since we assume the relation sizes and selectivities are unchanged, the total time complexity here remains $O(\frac{n^4}{b^3}log(\frac{n^4}{b^3}))$.
%

\ul{Join-Agg Operator:} The \textbf{number of vertices} in the data graph $|V|$ is $O(n)$ here because there will be two sets of multi-nodes here, one for $R_2$ and $R_3$ each. The \textbf{number of edges} in
the data graph $|E|$ is at most $O(\max(ab,n))$. Similarly as the above case, the
\textbf{traversal stage} will take $O(a*(|V|+|E|)) = O(\max(an, a^2b))$ time. However, the result generation stage is more complex here as we have to maintain ``paths'' at the reachable group
nodes and merge them at the end. Both the space and time complexity here is dominated by the number of different paths to the $g_i$ nodes. In the worst case, there may be $O(\frac{n^2}{b})$ such paths per $g_i$
node, giving us a total time complexity of $O(\frac{n^2}{b}\log(\frac{n^2}{b}))$ per source node. The overall complexity then is
$O(a\frac{n^2}{b}\log(\frac{n^2}{b}))$, and the space complexity is $O(a\frac{n^2}{b})$.

\ul{Pre-aggregation:} The partial pre-aggregation possibilities are somewhat limited here since the intermediate results contain a larger number of attributes, and thus have limited duplicity. If we
assume there is no reduction due to partial pre-aggregration, then the time complexity here is similar to the traditional approach, giving us $O(\frac{n^4}{b^3}log(\frac{n^4}{b^3}))$ time complexity.
However, another lower bound can be calculated using the similar worst-case assumption as above for the \joinAgg operator, where we assume all possible combinations of values exists in at least one join
result, giving us a time complexity of $O(a^4b \log(a^4b))$, with a space complexity of $O(a^4b)$.



\ul{Comparison:} The complexities of \joinAgg and partial pre-aggregation approaches are very different in this case. The partial pre-aggregation approach may perform somewhat better if the number of
unique values for the group attributes is small relative to the join attributes; however in that scenario, we expect the number of different paths to a $g_4$ node to be significantly smaller than $b^3$
(which assumes a worst-case situation that won't occur in practice). As above, we see that the Join-Agg space complexity is lower by a factor.

\topicul{Branching Join}
Next we consider a 5-relation branching query $Q = R_1(g_1,\_,j_1) \Join B(j_1, j_2, j_3, j_4) \Join R_2(j_2, \_, g_2) \Join R_3(j_3,\_, g_3)$ $ \Join R_4(j_4,\_, g_4)$, with a group by aggregate on four attributes from
four different relations.

%

\ul{Traditional RDBMS:} As above, the join computation time is dominated by the generation of the result tuples, giving us a total time of
$O(\frac{n^5}{b^4}log(\frac{n^5}{b^4}))$.

\ul{Join-Agg Operator:} The \textbf{number of vertices} in the data graph $|V| = 4a + 4b + n$ (since every tuple from $B$ will be a different node). The \textbf{number of edges} in the data graph $|E|$ is
at most $O(\max(n,ab))$. The \textbf{traversal stage} would again
take $O(a*(|V|+|E|)) = O(\max(an, a^2b))$ in total. The \textbf{result generation} however requires merging the lists of paths at each of the reachable grouping attribute nodes ($g_2, g_3, g_4$). It is easy to see that
maximum number of different paths from a given source node to any of the destination grouping nodes (say a $g_2$ node) is $n$, thus giving us a result generation time of $O(n\log(n))$ per source node. Since this has to
be done for each of the $g_1$ nodes, the total time for result generation is bounded by $O(an\log(n))$. Thus the overall complexity is $O(\max(a^2b, an\log(n)))$, with a space complexity of $O(\max(n, ab))$. Unlike the bounds so far, we don't attempt to substitute $n$ with $a$ and $b$ as the bounds become very loose in that case.

\ul{Pre-aggregation:} The partial pre-aggregation possibilities are somewhat limited here (outside of the input relations) since the intermediate results contain a larger number of attributes, and thus have limited duplicity. The
largest intermediate result we may generate here is $I(g_1, g_2, g_3, j_3, j_4)$, assuming we join $R_1$ with $B$ followed by $R_2, R_3, R_4$ in that order (with aggressive pre-aggregation at every step).
The size of $R_1 \Join B \Join R_2 \Join R_3$ can be estimated at $O(\frac{n^4}{b^3})$, and $I$ is the result of projecting out $j_1$ and $j_2$ from that join result (and any other attributes from those
        relations that did not participate in the join). However, it is difficult to estimate
the reduction in size from that projection. If $b$ is sufficiently large compared to $n$ (i.e., $b > \sqrt{n}$), then under uniformity assumptions, we expect minimal reduction in the size. Thus,
in general, we expect the total time and space complexity of the pre-aggregation approach to be very high compared to the \joinAgg operator.

\ul{Comparison:} Join queries with branching really illustrate the benefits of a holistic approach to executing such queries. The benefits over the traditional approach, even with aggressive
pre-aggregation, come from the ability to avoid generating large intermediate results, and exploit ``caching effects''.

\begin{figure*}[t]
\includegraphics[width=\textwidth]{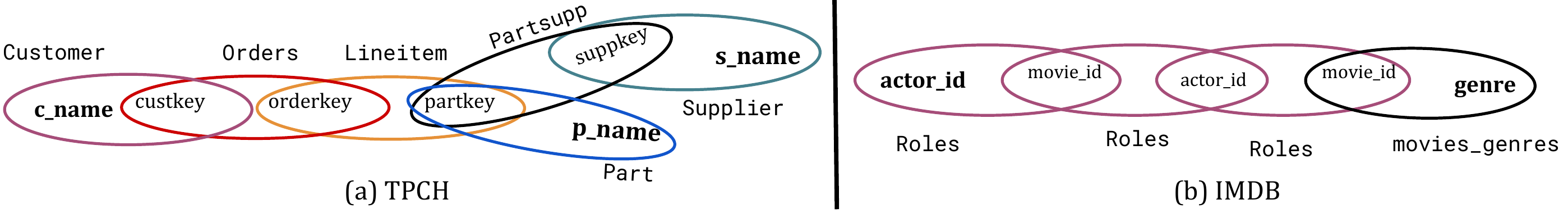}
\centering
\caption{Hypergraphs of real world queries in the experiments}
\label{fig:query-hypergraphs}
\end{figure*}

\section{Implementation Details}
\label{sec:implementation-details}
The data graph we load into memory is stored in a data structure resembling a Compressed Sparse Row (CSR) graph representation. We store a list of all \texttt{Edge} objects in the graph called \texttt{outNeighbors}, and a list of \texttt{Node} objects. Each \texttt{Node} object contains the \texttt{List} of attribute values the node is comprised of (note that a \texttt{Node} can have values from multiple attributes in its relation if it is a \textit{multi-node}). \texttt{Node}s also include one \texttt{Integer} for the \textit{offset} value, and one \texttt{Integer} that stores the \textit{number of neighbors} that particular node has.
The offset value points to the outgoing edges that correspond to the particular \texttt{Node}; i.e. the outgoing edges of a node \texttt{n} would start at \texttt{outNeighbors[n.offset]} and end at \texttt{outNeighbors[n.offset+ n.numNeighbors]}.
\texttt{Node}s also include an \texttt{Integer} value representing the \textit{type} of node (source, branching, group etc.). Group nodes in particular, are assigned a unique \texttt{Integer} that references the relation they came from in this field.
\texttt{Edge} objects store a reference to their outgoing neighbor \texttt{Node}, and a single \texttt{Integer} value for their \textit{multiplicity}. Path-ids are an integral part of the algorithm and are also stored as explicit objects, containing a \texttt{List} of branching node values, as well as an \texttt{Integer} for the path-id count.

\topicul{Stage 1: Data Graph Loading} During the loading for the data graph, each relation is \textit{sorted} by PostgreSQL, read in using a \texttt{JDBC} connector as a \texttt{List} of tuples, and each tuple is partitioned into its two $(x_l,x_r)$ subsets. Duplicate tuples (after projection) in each relation are also \textit{pre-aggregated} by PostgreSQL itself before being loaded into the data graph. A \texttt{HashMap} index is used to keep track of \texttt{Node} objects already loaded and access them in order to incrementally add each additional \texttt{Edge} to the data graph. For each relation being loaded, all children relations in the decomposition tree are subsequently loaded, as well as \texttt{Edge}s between \texttt{Node}s with overlapping values; these \texttt{Node}s intuitively map to joining tuples between the original relations. The CSR representation we use for our graph data structure is generally \textit{immutable}, we therefore make sure to properly load in each \texttt{Node} and all of its \texttt{Edge}s entirely before moving on to the next one so as to never require to shift anything in the \texttt{outNeighbors} list.

\topicul{Stage 2: Traversal} During the second step of the algorithm, the data graph is traversed in a \textit{dfs} fashion starting from each source node. The \texttt{visit()} method is recursively called over the neighbors of the current source node, properly propagating the multiplicity as well as the path-id along the way. We keep track of the path-ids in each iteration inside of a \texttt{HashMap}, therefore a single hash-lookup is required to check if the current path-id has already been visited by this current traversal. If so, we simply need to update its path-id count and continue with the next neighbor without continuing the traversal beyond that path since it has already been explored (for the current source node). This caching effect is one of the crucial optimizations that sets \joinAgg apart from other approaches such as \textit{partial pre-aggregation}~\cite{larson2002data}.


\topicul{Stage 3: Result Generation} After a full traversal of the graph starting from a single source node concludes, we now have enough information to output all the groups that contain that source node as a value. First we separate the set of \texttt{reached nodes} into \textit{buckets}, based on their \textit{type}. If and only if \textit{at least one} node from every relation in $G$ was touched, do we take \textit{any} further action in this stage.

Next we merge every c-pair in every node in each bucket into a single list of tuples ordered by path-id. We use a \textit{k-way merge} algorithm to do this since c-pairs are all naturally sorted by path-id at the end of Stage 2. Next, for every list $F_i$ generated by the previous step, we conduct a sort-merge join starting from the \textit{smallest list} that contains path-ids of the longest size. We therefore sort the $F_i$ lists first by path-id length (in a \textit{decreasing} order), and then sort them by list size (in an \textit{increasing} order).
After the sort-merge join is completed, the result is sorted by the value of each output group lexicographically.

\subsection{Pre-aggregation Implementation}
In order to experimentally support our hypothesis described in Section~\ref{sec:complexity}, we implemented a simple in-memory database in Java which allowed us to manually describe query plans. We stored in-memory rows as Java \texttt{LinkedList}s, and stored all values as \texttt{String} objects, as we did in the \joinAgg implementation, for the sake of consistency. We implemented a \textit{hash-join} over two sets of tuples, \textit{project} over a set of tuples, as well as a \textit{hash-aggregate} group by operation over a set of tuples. We use the standard algorithms for hash-join. In particular, we create a \texttt{HashMap} on the join condition value for every tuple $s_i$ in the \textit{smallest} of the two sets of tuples, and probe that \texttt{HashMap} for every tuple $l$ in the larger set, to generate all combinations $l$,$s_i$.

\topicul{Optimizations} We included a few optimizations in order for our code to be as comparable as possible to a real in-database implementation. Firstly, we combined the project and hash-aggregate operators so that tuples are only read once, unnecessary columns are projected out, and the column is then hashed for aggregation in the same step. Moreover, due to the fact that each tuple's values are static (before it is joined), we compute the \texttt{hashCode()} of every tuple only once, upon its creation so that it doesn't need to be computed again when hashing the tuple (either at the join or aggregation step). At the hash-join stage, we allocate new memory for the output tuples only when outputting the join result. We store the pre-aggregated count at every stage in a separate field for each tuple.

\section{Experimental Evaluation}
\label{sec:experiments}

We present an experimental evaluation over a series of synthetic and real datasets that showcase the benefits and trade-offs of the \joinAgg operator. We've generated $3$ synthetic datasets for three
types of queries described in Section~\ref{sec:complexity}, the hypergraphs for which can be seen in Figure~\ref{fig:example-hypergraphs}.
We also present experiments on queries over TPCH~\cite{tpch-dataset} (using scale factor \texttt{SF=1}), DBLP~\cite{dblp-dataset},  ORDS~\cite{ordsDataset} and IMDB~\cite{relational-repo}.
Each dataset is associated with a specific query, query hypergraphs for which are shown in Figure~\ref{fig:query-hypergraphs}. Datasets DBLP and ORDS are both simple self-joins.
Additional information about these datasets is shown in Table~\ref{tab:datasets}.

We implemented a prototype of the \joinAgg operator entirely in \textit{Java}. We load the data directly from PostgreSQL into the JVM by using the \texttt{JDBC} connector. Our aim with this prototype is to showcase that the execution of aggregate queries over large-output joins can, in many situations, be evaluated \textit{more efficiently} even \textit{outside} of the RDBMS including the often substantial \textit{overhead} of loading the data from PostgreSQL into the JVM. We advocate that a native version of \joinAgg implemented natively within an RDBMS itself in a lower level language would demonstrate an even wider performance gap in favor of \joinAgg. The main reason is that loading the data graph would naturally be significantly faster, because reading the data tuple-at-a-time using \texttt{JDBC} is a significant portion of the loading time overhead.

These experiments were all done on a single machine running Red Hat Enterprise Linux Server 6.9, with 64GB of RAM, and an Intel(R) Xeon(R) CPU E5-2430 0 @ 2.20GHz, using PostgreSQL version 9.4.18 and Java 8.

{
\begin{table}
\center
    \small
    \begin{filecontents*}{numbers/synthetic-datasets.csv}
Query, Dataset,Selectivity,JoinR,Groups,Load(s)
Self-Join,S1,0.001     ,500 M  ,6.25 M  ,11.263
         ,S2,0.003     ,167 M  ,6.25 M  ,12.729
         ,S3,0.1       ,5.5 M  ,3.43 M  ,29.182
Chain    ,C1,0.1       ,837 M  ,5.04 M  ,15.203
         ,C2,0.3       ,64 M   ,1.71 M  ,20.198
         ,C3,0.5       ,23 M   ,1.04 M  ,22.048
Branch   ,B1,0.001/0.8 ,1.4 B  ,0.12 M  ,35.935
         ,B2,0.1/0.1   ,549 M  ,0.12 M  ,44.781
         ,B3,0.3/0.5   ,9.9 M  ,9.76 M  ,32.804
         ,      ,      ,       ,           ,
Real     ,TPCH  ,      ,24 M   ,23.9 M  ,161.197
Queries  ,DBLP  ,      ,105 M  ,87.8 M   ,253.536
         ,ORDS  ,      ,59 M   ,7.50 M    ,20.881
         ,IMDB  , ,4.4 B       ,13.1 M   ,138.956
\end{filecontents*}
\csvautotabular{numbers/synthetic-datasets.csv}
    \caption{Characteristics about all synthetic and real datasets used in the experiments. \textit{JoinR} shows the size of the join result before aggregation in Million (M) or Billion (B) tuples. \textit{Groups} shows the number of groups output for each query in each dataset. \textit{Load} is the total time required (in seconds) to load the data from PostreSQL to the in-memory data graph. }
    \label{tab:datasets}
\end{table}
}
{
\begin{table}
\center
    \small
    \begin{filecontents*}{numbers/sample-datasets.csv}
Dataset, Max Interm. Result, JoinAgg Mem, PreAgg Mem
P1, \num[group-separator={,}]{987285}  , 0.097 ,  0.259
P2, \num[group-separator={,}]{3755151} , 0.233 ,  1.19
P3, \num[group-separator={,}]{13414963}, 0.547 ,  3.3
P4, \num[group-separator={,}]{27952709}, 0.976 ,  6.6
P5, \num[group-separator={,}]{45762103}, 1.1   ,  $>$9GB
P6, \num[group-separator={,}]{66326006}, 1.3   ,  $>$9GB
\end{filecontents*}
\csvautotabular{numbers/sample-datasets.csv}
    \caption{Samples from the B2 dataset, the max memory consumption (max heap used in GB) when running \joinAgg or pre-aggregation respectively, as well as the size of the max intermediate result (in rows) that needed to be processed when using pre-agg.}
    \label{tab:sample-datasets}
\end{table}
}

\subsection{Synthetic Datasets}
The synthetic datasets that were used for studying the behavior on the example queries showcased in Section~\ref{sec:complexity}, were generated by pulling from a uniform distribution (using Java's \texttt{Random} class) of a certain set of values, based on the \textit{selectivity} we wanted to emulate each time. We define the term \textit{selectivity} as $s = |\pi_j(R)|/|R|$, where $\pi_j(R)$ the domain of unique values of attribute $j$ in relation $R$. For each \texttt{S1,S2,S3} dataset, we generate a single relation $R(g,j)$ for which the join selectivity when joining with itself is roughly the one reported in Table~\ref{tab:datasets}. Similarly, for each \texttt{C1,C2,C3} dataset, we again generate a single relation with the specified join selectivity and use copies of that relation for each part of the chain--therefore all joins in the chain portray the same selectivity.
For the \texttt{B1,B2,B3} datasets, there are two different selectivities specified, the first is for the join $R_1(g1,j) \Join R_2(j,b)$ and the second for the joins $R_2(j,b) \Join R_3(b,g_2)$ and $R_2(j,b) \Join R_4(b,g_3)$. Again, for each of the join condition attributes in each table, we generated each tuple by drawing from a uniform distribution of integers in the range $[0,s*|R|]$. Group attributes were generated the exact same way. The range that we used for generating the group attribute in each of these relations is roughly reflected by the number of output groups generated by the queries. For the sake of simplicity all generated relations are of size $|R_i| = 500,000$ tuples.

\begin{table}[h!]%
  \centering
  \begin{tabular}{clcc}
  \hline
  Dataset       & S1          & S2          & S3        \\
  (groups/size) & (6.25 M/80) & (6.25 M/26) & (3.4 M/1) \\
                &             &             &           \\ \hline
  PostgreSQL    & 499 s       & 181 s       & 11 s      \\
  JOIN-AGG      & 38 s        & 28 s        & 33 s      \\ \hline
  \end{tabular}
\caption{Experiment for the Self-join example}
\label{tab:self-join-runs} %
\end{table}

\begin{table}[h!]%
  \centering
  \begin{tabular}{clcc}
\hline
Dataset       & C1        & C2         & C3       \\
(groups/size) & (5 M/165) & (1.7 M/37) & (1 M/22) \\
              &           &            &          \\ \hline
PostgreSQL    & 512 s     & 65 s       & 18 s     \\
JOIN-AGG      & 21 s      & 22 s       & 24 s     \\ \hline
\end{tabular}
\caption{Experiment for the Chain example}
\label{tab:chain-runs} %
\end{table}

\begin{table}[h!]%
  \centering
  \begin{tabular}{clcc}
  \hline
  Dataset       & B1           & B2          & B3        \\
  (groups/size) & (125 K/11 K) & (125 K/4 K) & (976 K/1) \\
                &              &             &           \\ \hline
  PostgreSQL    & 1104 s        & 393 s       & 18 s      \\
  JOIN-AGG      & 136 s         & 226 s       & 55 s      \\ \hline
  \end{tabular}
\caption{Experiment for the Branching example}
\label{tab:branch-runs} %
\end{table}

\begin{table}[h!]%
  \centering
  \begin{tabular}{clccc}
  \hline
  Dataset        & TPCH     & DBLP     & ORDS       &  IMDB  \\
  (groups/size)  & (23 M/1) & (87 M/1) & (7.5 M/7)  &  (13 M/340)  \\
                 &          &          &            &              \\ \hline
  PostgreSQL     & 71 s     & 172 s    & 95 s       &  3422 s      \\
  JOIN-AGG       & 248 s    & 384 s    & 31 s       &  1156 s       \\ \hline
  \end{tabular}
\caption{Experiment for queries over real datasets}
\label{tab:real-query-runs} %
\end{table}

\begin{figure}[t]
\includegraphics[width=0.50\textwidth]{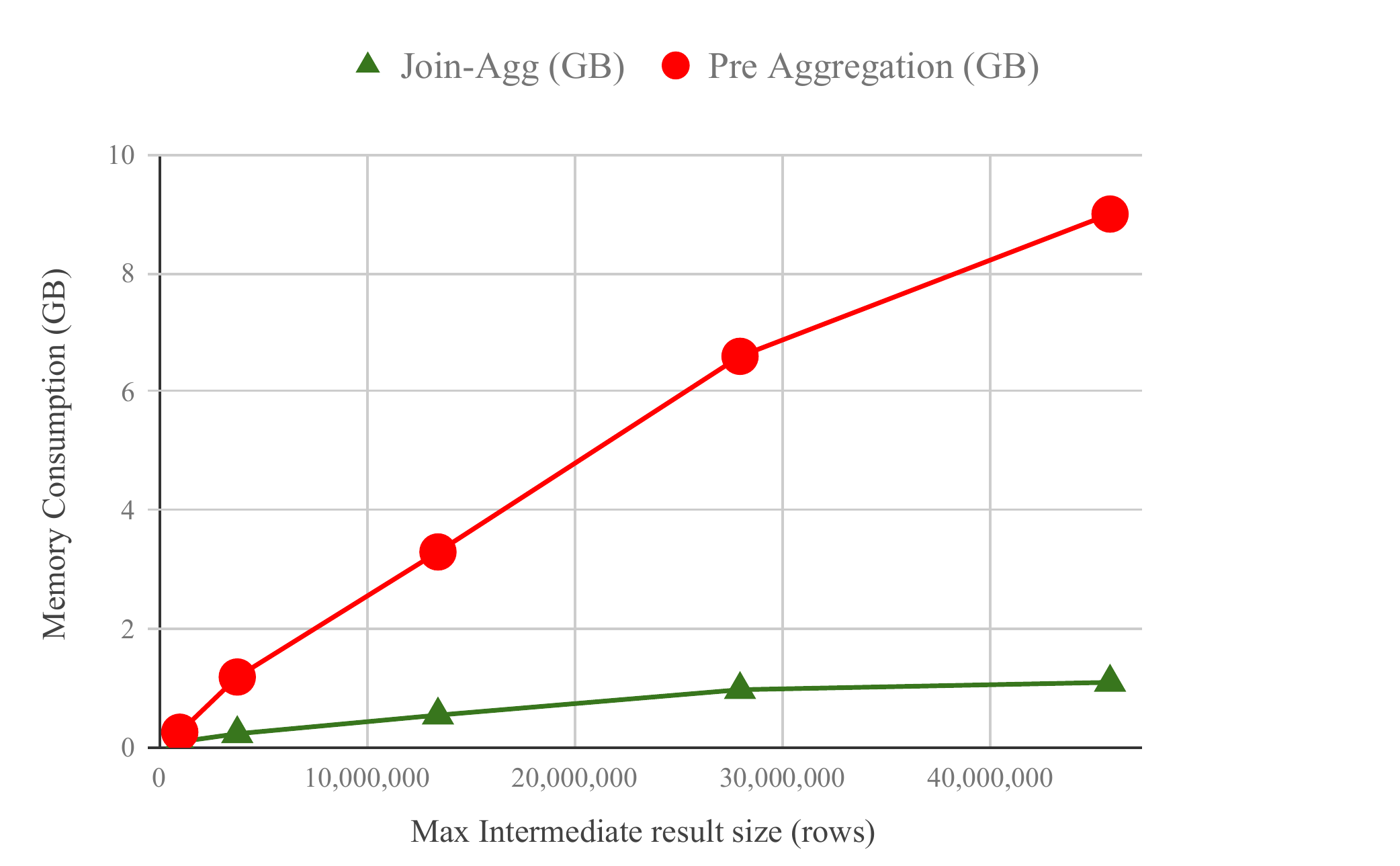}
\centering
\caption{Maximum memory consumption (max heap used), at any point during execution. Each value in the y-axis represents the largest intermediate result we needed to store when using pre-aggregation at every stage of the join}
\label{fig:memory-ja-preagg}
\end{figure}

\begin{figure}[t]
\includegraphics[width=0.40\textwidth]{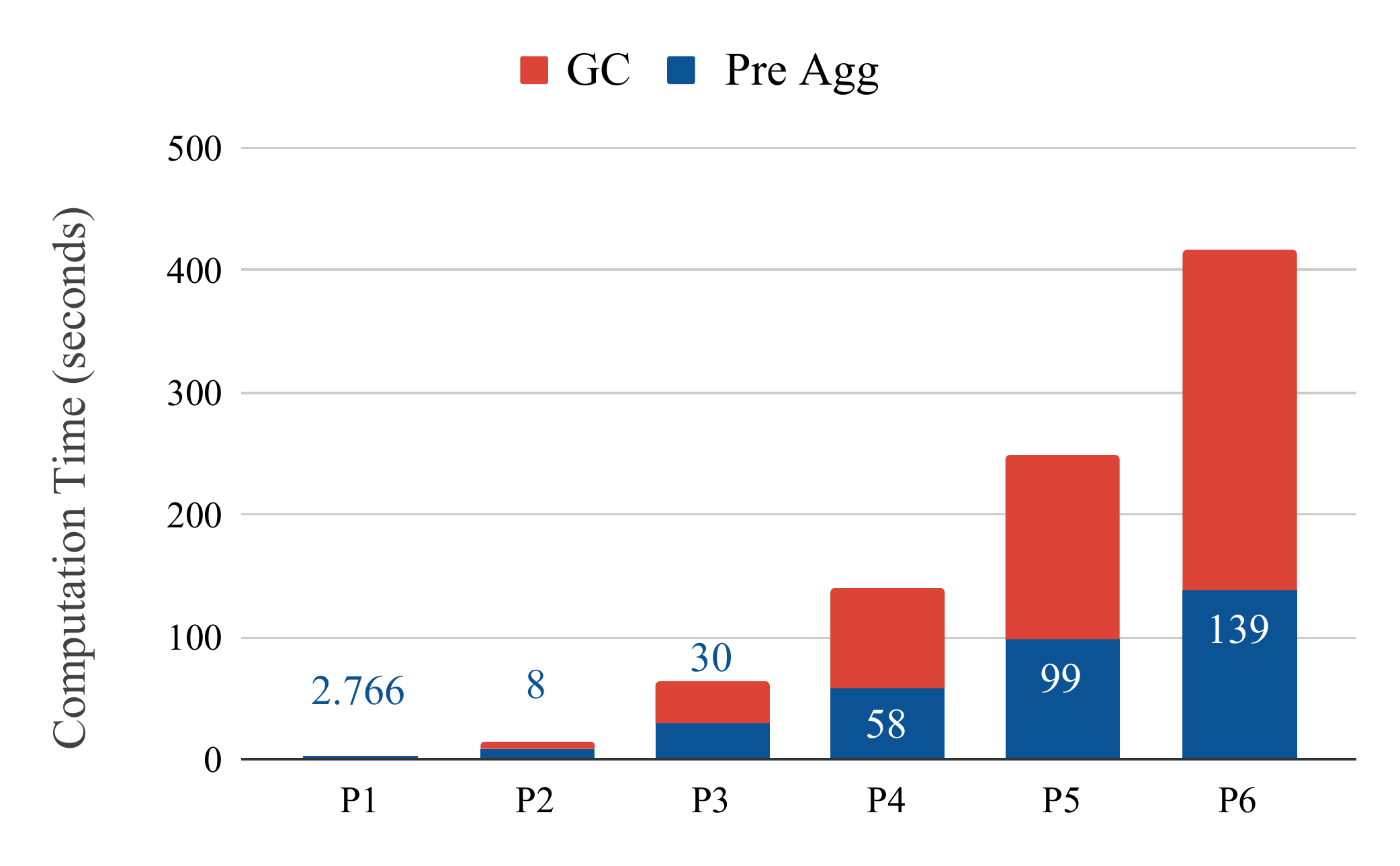}
\centering
\caption{Total computation time spent when using pre-aggregation per sample, showing the portion of the computation time spent on garbage collection (GC)}
\label{fig:preagg}
\end{figure}

\begin{figure}[t]
\includegraphics[width=0.40\textwidth]{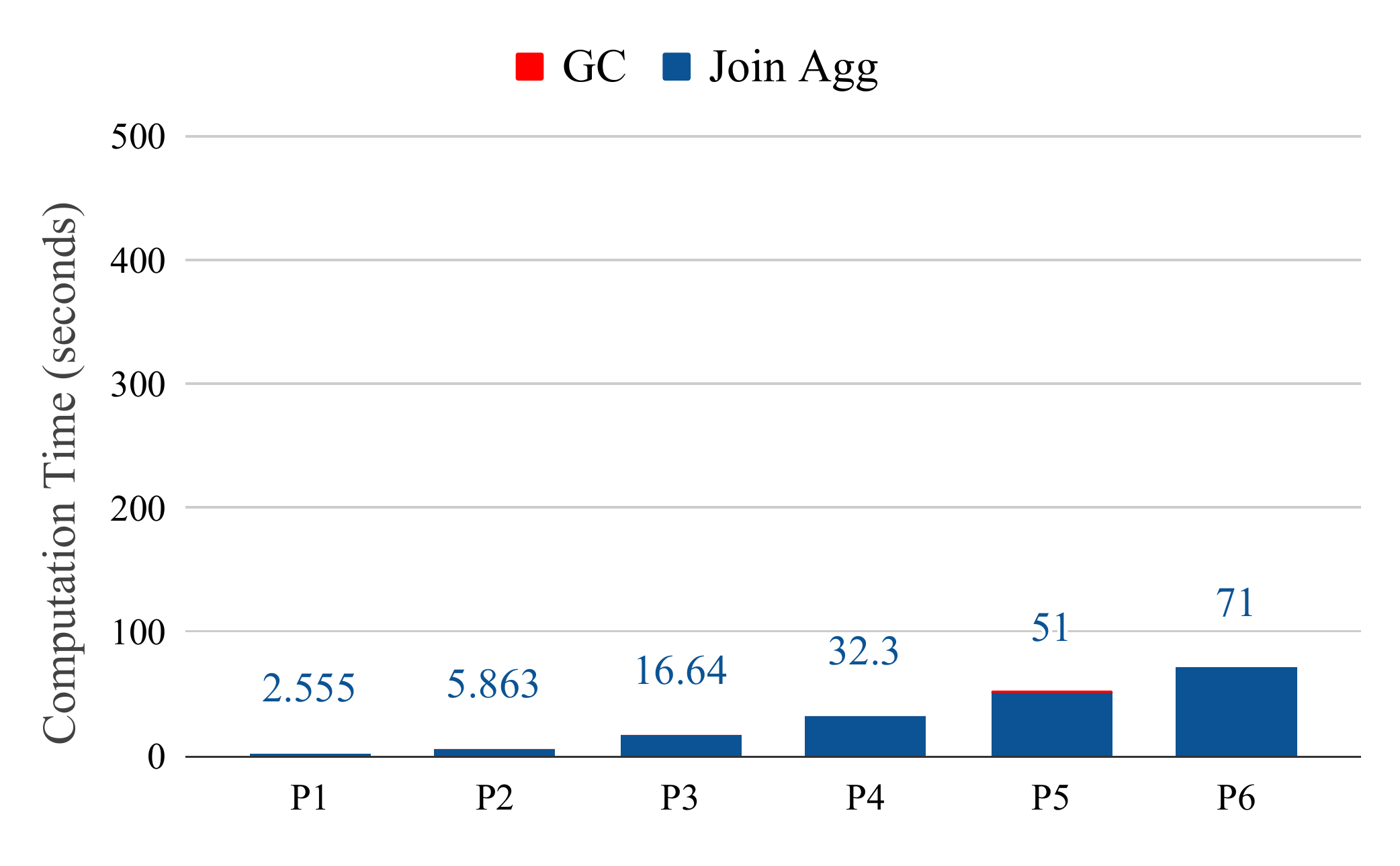}
\centering
\caption{Total computation time spent when using join-agg per sample, showing the portion of the computation time spent on garbage collection (GC)}
\label{fig:joinagg}
\end{figure}

\begin{figure}[t]
\includegraphics[width=0.40\textwidth]{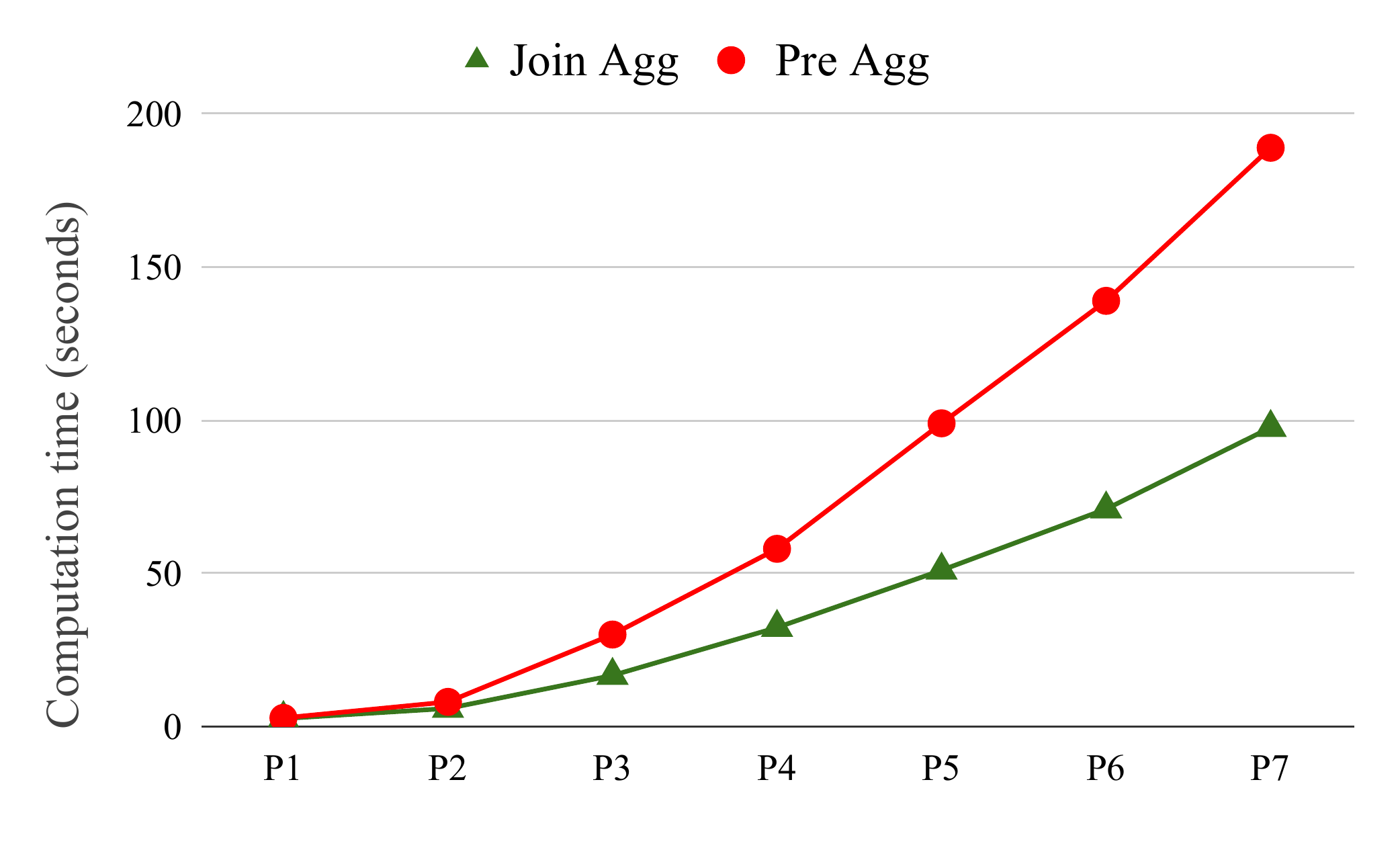}
\centering
\caption{Only computation time (excluding GC time) for every sample dataset.}
\label{fig:gap}
\end{figure}


\subsection{Tuning PostgreSQL}
We evaluate the performance of \joinAgg by comparing it to running these queries directly over a state of the art RDBMS; PostgreSQL. One of the database parameters that proved crucial for these queries is \texttt{work\_mem}, which specifies the amount of memory every distinct query operator can utilize within a single query. In a data warehouse setting, given the specifications of the server machine we used, \texttt{work\_mem} would normally be set to around \texttt{256MB}.  Setting \texttt{work\_mem} to a very high value is generally not recommended because it increases the risk of the database running out of usable memory very quickly as \textit{multiple} user queries are executed simultaneously. \joinAgg on the other hand only asymptotically needs as much memory as is required to store the data graph, per query, thus enabling multiple such queries to practically be run simultaneously and efficiently whereas PostreSQL would need to use slow methods (e.g. use \texttt{SortMerge} Joins and \texttt{GroupAggregate} for aggregation).


Nevertheless, to showcase the \textit{best possible} performance we could get out of PostgreSQL on this specific machine, we set \texttt{work\_mem} to \texttt{10GB}. This essentially allowed the PostgreSQL query planner to mostly choose the \texttt{HashAggregate} operator instead of \texttt{GroupAggregate} which can be orders of magnitude slower, depending on whether the \texttt{Sort} phase happens in memory or on disk. The query plan generated by PostgreSQL when running these aggregate queries, showed that it always chooses to use \texttt{SortMerge} Joins, and \texttt{GroupAggregate}, when it estimates the value of \texttt{work\_mem} isn't high enough to fit the hash-table based on the estimated number of output groups.

We also observe that PostgreSQL is completely unable to estimate the number of tuples in the result set, and uses the \textit{same cardinality estimate} as the result of the join, for estimating the number of groups in the result. Anecdotally, we estimate this is as the primary reason PostgreSQL may choose to use \texttt{GroupAggregate} and \texttt{SortMerge} joins, to ensure that the query will not run out of memory instead of trying to use operators that require hashing, which are faster but require significantly larger amounts of memory.

\subsection{Join-Agg Performance Analysis}
We studied the three basic types of queries that constitute the baseline for most join-aggregation queries over a database. Our overall conclusion was that \joinAgg can make a huge difference in query execution time for a query, as that query outputs \textit{larger groups}. The larger the size of the groups in the output, the more there is to gain from \joinAgg. In cases where the output is comprised of small groups (i.e. of size 1), \joinAgg portrays comparable performance to the traditional approach when taking into account the fact that a large portion of the execution time in \joinAgg is taken \textit{loading} the data out of the database.

Table~\ref{tab:self-join-runs}, showcases the performance of a join-aggregation query over a single \textit{self-join}. We can see that \joinAgg performs over to an order of magnitude better than PostgreSQL when we have a relatively large group size and the gap between the two systems closes as that average size leans towards $1$. This makes sense since outputting many groups of size $1$ indicates the intermediate result is close in size to the final result, thus materializing it is mostly inevitable. Similar behavior is seen for the \textit{chain} example shown in Table~\ref{tab:chain-runs}. Note that when we have multiple non-key joins in a row as is the case with this example, the selectivities of those joins don't need to be absurdly low for \joinAgg to have a substantial difference in performance. This is because the intermediate results keep expanding as non-key joins progress resulting in the output of a very large set of tuples that then need to be aggregated.


In Table~\ref{tab:branch-runs} we can see that for datasets \texttt{B1,B2,B3}, where we have three group attributes from different relations, showcase a similar performance trend as the other examples. In the datasets that output large groups, \joinAgg performs up to an order of magnitude better whereas the performance of \texttt{B3} which outputs groups of size $1$ on average, is comparable to PostgreSQL. Particularly for dataset \texttt{B1}, we have a very low selectivity ($0.001$) join for $R_1 \Join R_2$ whereas the other joins portray a high selectivity ($0.8$). We can see that even a single low-selectivity join in this complex query, can result in a huge ($1.4 B$ tuples) intermediate output which \joinAgg helps to avoid materializing.


In the real datasets we experimented with, showcased in Table~\ref{tab:real-query-runs}, we observe results consistent with the synthetic datasets. The \texttt{DBLP} (rf. Figure~\ref{fig:example-hypergraphs}a) and \texttt{TPCH} (rf. Figure~\ref{fig:query-hypergraphs}a) queries output very small groups, causing the time for loading the data graph to dominate the computation. The dataset \texttt{ORDS}~\cite{ordsDataset} is a typical \textit{market basket} dataset of invoices that contain multiple items.
We are querying all item pairs and counting \textit{how many times} they were bought together. \texttt{IMDB}, is graph pattern counting query over the \textit{IMDB} movie graph as seen in Figure~\ref{fig:query-hypergraphs}b. This query counts the number of paths between an actor and a genre, two hops away from that actor, i.e. even genres of movies that co-actors of theirs played in. For both of the latter queries the groups portray a higher average size and the benefits of \joinAgg start becoming apparent.

\subsection{Pre-aggregation Performance Analysis}

To experimentally validate our hypothesis in regards to how using pre-aggregation stacks up against our approach, we sampled the \texttt{B2} dataset -- incrementally taking a larger sample. Information about the samples can be seen in Table~\ref{tab:sample-datasets}.

Figure~\ref{fig:memory-ja-preagg} showcases the difference in memory requirements between \joinAgg and pre-aggregation. We can see that in the case of pre-aggregation, as the size of the largest intermediate result required for the query \textit{after} using aggressive pre-aggregation at every stage of the join increases, the maximum amount of memory required to complete the query increases rapidly. The memory required when it comes to executing \joinAgg however increases slowly since it only has to do with the size of the input data in combination with the largest amount of c-pairs that need to be stored at a single iteration (after we process any one source node).

Figures~\ref{fig:preagg} and \ref{fig:joinagg} showcase the computation time required for the execution of the branching query shown in Figure~\ref{fig:example-hypergraphs}c. Due to the fact that our pre-aggregation implementation is relatively simple and done in Java (as discussed in detail in Section~\ref{sec:implementation-details}), a large portion of the computation comes down to garbage collection time. If however we only look at the amount of time spent doing actual computation as shown in Figure~\ref{fig:gap}, we can clearly see the gap in performance between the two techniques, as was expected based on the complexity analysis in Section~\ref{sec:complexity}.

\section{Related Work}
\label{sec:related-work}
Here we discuss the closely related work on executing these types of queries in different contexts within RDBMSs.

\topic{Factorized Databases and Worst-Case Optimal Joins}
The data-graph paradeigm that we propose in this paper is reminiscent of the factorized representation of conjunctive query results previously proposed by Olteanu et al.~\cite{olteanu2012factorised}. Both representations aim at representing the underlying join while reducing the amount of data stored in order to do so. The data graph can also be connected to the idea of a \textit{tuple hypergraph} which can cover all tuples in a query result ~\cite{kara2018covers}; it however serves a very different purpose.

Our main objective in this work is to be able to compute \textit{aggregations} over a representation like the data graph, especially in the case of complex acyclic joins.

Several different works have considered the problem of executing group by aggregate queries against a factorized representation of a conjunctive
query~\cite{bakibayev2013aggregation,khamis2018ac,khamis2018functional,khamis2019boolean,schleich2019layered,schleich2016learning,olteanu2016factorized}. The key guarantees like constant-delay enumeration, however, do
not extend to the kind of group by queries we focus on in this work, e.g., the ``branching'' query $R_1(g_1,j), R_2(j,b), R_3(b,g_  3), R_4(b, g_2)$. Because all of $g_1,
g_2, g_3$ (group by attributes) need to be present in the output, either (a) one of the other attributes needs to be eliminated (which requires generation of a large
intermediate result), or (b) we have to iterate over all combinations of values for $g_1, g_2, g_3$ and compute the aggregate value for each combination (which can be
prohibitively expensive if either the sizes or the number of group by attributes is large). Our work here, thus, can be seen as exploring an alternative approach to computing aggregates over the factorized representation.

As we discussed earlier, recent work on worst-case optimal joins~\cite{ngo2012worst,DBLP:conf/icdt/Veldhuizen14,koutris2016worst,ngo2014skew} shows how to avoid large intermediate results during execution of multi-way join queries;
Joglekar et al.,~\cite{joglekar2015aggregations,joglekar2016ajar} discuss how to generalize that to aggregate queries. Their approach is largely complementary to ~\cite{bakibayev2013aggregation}, as well as our line of work.
Recent work on FAQ~\cite{abo2016faq} proposed a generalized way of viewing a very common type of aggregation query called a Functional Aggregate Query which they see parallels in multiple scenarios other than databases e.g. matrix multiplication, probabilistic graphical models, and logic. The InsideOut algorithm proposed in FAQ however is not focused on executing SQL queries, like our work as well as the factorized databases work is aimed at doing. FAQ also assumes an optimal variable order, while this paper does not explore the benefits of choosing the optimal variable order (tree decomposition in our case).

\topic{Iceberg Queries} An \textit{iceberg query} is a particular class of SQL queries, defined as an aggregate query, counting occurrences of target group instances of the \texttt{GROUP BY} clause columns, and filtering the results post-aggregation using a \texttt{HAVING} clause. These queries typically return a small fraction of the overall (potentially large) join result, (the tip of the iceberg). Iceberg queries are clearly a special case of the queries we're studying in this paper.

Fang et al.~\cite{fang1999computing} propose a wide array of techniques for computing iceberg queries which focus on \textit{minimizing the passes} done over the data (Disk I/O), being able to answer such queries in reasonable time, and doing so with a small amount of memory.
The authors focus on combining two techniques: coarse counting (probabilistic counting), and sampling. These techniques may start causing issues as the final result increases in size. In a similar setting there has been work on efficiently computing the \textit{iceberg} \texttt{CUBE}~\cite{beyer1999bottom,han2001efficient,xin2003star}, which is largely orthogonal to ours, since this paper focuses on the general case of outputting all groups. Developing techniques for more efficient \textit{iceberg} queries using our \joinAgg operator are delegated to future work.

Walenz et al.~\cite{walenz2017optimizing}, presented a series of optimizations applicable to certain types of iceberg queries. The main focus of this work is to use formal methods towards
\textit{automatically identifying} whether a general SQL query would benefit from certain specialized optimizations for evaluating certain types of iceberg queries, as well as towards automatically using
such optimizations during evaluation. The optimizations they consider involve pruning techniques based on memoization and complex non-equality join conditions. Given a general SQL query, their methods
systematically identify if any optimization technique is applicable, and use it during execution of the query. Similarly to us, the authors implement and wrap the above optimizations into a custom database \textit{join operator}. The work in this paper is largely orthogonal to ours since it mainly deals with complex join conditions, it does not focus on minimizing extra memory consumption during execution, and is more aimed at providing formal methods for automatically identifying queries that would benefit from these specialized optimizations.

\topic{Similarity Joins} Work on \textit{similarity joins}~\cite{jiang2014string,xiao2011efficient,wang2010trie,li2011pass} uses various techniques to prune join computation. In a similarity join
between two relations, (on a \textit{string} join condition), a pair of tuples join if their join attribute similarity surpasses a threshold. This can be directly mapped onto the iceberg query problem
where the aggregation function is \texttt{COUNT}. From this perspective, iceberg queries aim at finding the tuples in the result that have a certain number of join condition attributes in common, which
surpasses a threshold. Similarity join techniques are almost exclusively signature-based (strings are collapsed into smaller signature sets). In a lot of these approaches, an ``inverted index'' is built
beforehand, which in a sense resembles our in-memory graph structure. These join algorithms are however only studied for binary operations, similar to the self-join case.


\topic{Data Reduction Operators} Work on lazy vs eager aggregation~\cite{yan1995interchanging,yan1995eager,yan1994performing} aims at re-arranging group-by operators in the logical query plan tree, moving them after or before joins accordingly. These techniques don't deal with avoiding materialization of intermediate results in situations when group by operators cannot be pushed down. Aggregation can only be pushed down if it can be partially applied to a single relation, thus reducing that relation's cardinality. In the general case however when the query contains a series of group by attributes, each one coming from a different relation, there's no way to apply any complex aggregation to a single relation because the aggregation applies to the result.

Larson et al., describe techniques for doing \textit{partial pre-aggregation}~\cite{larson2002data}. They describe a way to apply pre-aggregation to input relations when another aggregation is conducted
on their join result. A simple hash table is used to aggregate tuples in the same relation, thus reducing the number of tuples joining with the next relation. As groups are pre-aggregated sequentially, if the number of pre-aggregated groups surpasses the memory capacity,  partially pre-aggregated tuples are output to make room for new groups; therefore the pre-aggregation can be incomplete. Those same-group tuples will be aggregated later on at the final aggregation step. They also describe techniques to combine this pre-aggregation process with a join by pre-aggregating while reading the relation and joining the output partially pre-aggregated tuples with the tuples from the inner relation. These techniques however apply to a single binary join at a time, and as we show in Figure~\ref{fig:memory-ja-preagg}, Join-Agg provides substantial memory benefits than partial pre-aggregation especially when the two are combined and we use pre-aggregation before loading in the data graph.

As previously mentioned, the way we load the data graph into memory is reminiscent of these data reduction operators since we are pre-aggregating all relations
to compute the multiplicity of each edge in the data graph. The creation of \textit{multi-nodes} in the data graph can also be seen as an even more \textit{effective}
form of pre-aggregation. For example in Figure~\ref{fig:running-ex}, looking at relation $B$, we can see that any standard pre-aggregation operator would reduce the
relation to at least $2$ tuples with \texttt{jc1,jd1} appearing twice whereas we load a single \texttt{jc1,jd1} node. Our techniques are comparable with partial
pre-aggregation in the case where there are no \textit{branching} relations. As branching relations and multiple group by attributes are included in a complex join, our
technique enables computation caching at the level of path-ids which can reduce the number of paths taken in the data graph during Stage 2 of the algorithm. The partial
pre-aggregation technique has no means of skipping these paths and may require computing all of the joins associated with those paths potentially multiple times.

\section{Conclusion}
\label{sec:conclusion}
In this paper, we proposed a multi-way database operator called \joinAgg that enables the memory-efficient execution of aggregation queries over joins that output large
intermediate results, by executing the query over a \textit{graph representation} of the underlying data called the data graph. We presented a detailed complexity
analysis comparing our approach to the traditional binary joins-based approach as well as an idealized partial pre-aggregation approach. Our experiments show that \joinAgg
operator can be over an order of magnitude more efficient than the traditional approach for a wide variety of queries, even when implemented outside of the RDBMS. 
We advocate that multi-way database operators may be the answer to dealing with the ``non-normalized'' data of the real world, which often leads to expensive non-key joins along with other aforementioned issues. They will enable users to continue leveraging RDBMSs for their OLAP analyses, requiring smaller amounts of resources.


%
%
%

\bibliographystyle{abbrv}
\bibliography{aggregates}

\begin{thebibliography}{10}

\bibitem{tpch-dataset}
\url{http://www.tpc.org/tpch/}.

\bibitem{dblp-dataset}
Dblp dataset.
\newblock
  \url{https://dblp.uni-trier.de/faq/How+can+I+download+the+whole+dblp+dataset}.

\bibitem{aberger2015emptyheaded}
C.~R. Aberger, S.~Tu, K.~Olukotun, and C.~R{\'e}.
\newblock {E}mpty{H}eaded: A relational engine for graph processing.
\newblock In {\em SIGMOD}, 2016.

\bibitem{abo2016faq}
M.~Abo~Khamis, H.~Q. Ngo, and A.~Rudra.
\newblock {FAQ:} questions asked frequently.
\newblock In {\em PODS}, pages 13--28. ACM, 2016.

\bibitem{avnur2000eddies}
R.~Avnur and J.~M. Hellerstein.
\newblock Eddies: Continuously adaptive query processing.
\newblock In {\em SIGMOD}, 2000.

\bibitem{bakibayev2013aggregation}
N.~Bakibayev, T.~Ko{\v{c}}isk{\`y}, D.~Olteanu, and J.~Z{\'a}vodn{\`y}.
\newblock Aggregation and ordering in factorised databases.
\newblock {\em PVLDB}, 6(14), 2013.

\bibitem{beyer1999bottom}
K.~Beyer and R.~Ramakrishnan.
\newblock Bottom-up computation of sparse and iceberg cube.
\newblock In {\em ACM Sigmod Record}, pages 359--370, 1999.

\bibitem{deshpande2004initial}
A.~Deshpande.
\newblock An initial study of overheads of eddies.
\newblock {\em ACM SIGMOD Record}, 33(1):44--49, 2004.

\bibitem{fang1999computing}
M.~Fang, N.~Shivakumar, H.~Garcia-Molina, R.~Motwani, and J.~D. Ullman.
\newblock Computing iceberg queries efficiently.
\newblock In {\em VLDB}, 1999.

\bibitem{gottlob2016hypertree}
G.~Gottlob, G.~Greco, N.~Leone, and F.~Scarcello.
\newblock Hypertree decompositions: Questions and answers.
\newblock In {\em PODS}, 2016.

\bibitem{han2001efficient}
J.~Han, J.~Pei, G.~Dong, and K.~Wang.
\newblock Efficient computation of iceberg cubes with complex measures.
\newblock In {\em ACM SIGMOD Record}, 2001.

\bibitem{relational-repo}
O.~S. Jan~Motl.
\newblock Relational dataset repository.
\newblock \url{https://relational.fit.cvut.cz/}.

\bibitem{jiang2014string}
Y.~Jiang, G.~Li, J.~Feng, and W.-S. Li.
\newblock String similarity joins: An experimental evaluation.
\newblock {\em Proceedings of the VLDB Endowment}, 7(8):625--636, 2014.

\bibitem{joglekar2015aggregations}
M.~Joglekar, R.~Puttagunta, and C.~R{\'e}.
\newblock Aggregations over generalized hypertree decompositions.
\newblock {\em arXiv preprint arXiv:1508.07532}, 2015.

\bibitem{joglekar2016ajar}
M.~R. Joglekar, R.~Puttagunta, and C.~R{\'e}.
\newblock Ajar: Aggregations and joins over annotated relations.
\newblock In {\em PODS}, 2016.

\bibitem{kara2018covers}
A.~Kara and D.~Olteanu.
\newblock Covers of query results.
\newblock In {\em 21st International Conference on Database Theory}, 2018.

\bibitem{khamis2018functional}
M.~A. Khamis, R.~R. Curtin, B.~Moseley, H.~Q. Ngo, X.~Nguyen, D.~Olteanu, and
  M.~Schleich.
\newblock On functional aggregate queries with additive inequalities.
\newblock {\em arXiv preprint arXiv:1812.09526}, 2018.

\bibitem{khamis2018ac}
M.~A. Khamis, H.~Q. Ngo, X.~Nguyen, D.~Olteanu, and M.~Schleich.
\newblock Ac/dc: In-database learning thunderstruck.
\newblock In {\em Proceedings of the Second Workshop on Data Management for
  End-To-End Machine Learning}, page~8. ACM, 2018.

\bibitem{khamis2019boolean}
M.~A. Khamis, H.~Q. Ngo, D.~Olteanu, and D.~Suciu.
\newblock Boolean tensor decomposition for conjunctive queries with negation.
\newblock In {\em 22nd International Conference on Database Theory}, 2019.

\bibitem{koutris2016worst}
P.~Koutris, P.~Beame, and D.~Suciu.
\newblock Worst-case optimal algorithms for parallel query processing.
\newblock In {\em LIPIcs-Leibniz International Proceedings in Informatics},
  volume~48. Schloss Dagstuhl-Leibniz-Zentrum fuer Informatik, 2016.

\bibitem{larson2002data}
P.-A. Larson.
\newblock Data reduction by partial preaggregation.
\newblock In {\em ICDE}, 2002.

\bibitem{li2011pass}
G.~Li, D.~Deng, J.~Wang, and J.~Feng.
\newblock Pass-join: A partition-based method for similarity joins.
\newblock {\em Proceedings of the VLDB Endowment}, 5(3):253--264, 2011.

\bibitem{ngo2012worst}
H.~Q. Ngo, E.~Porat, C.~R{\'e}, and A.~Rudra.
\newblock Worst-case optimal join algorithms:[extended abstract].
\newblock In {\em PODS}, 2012.

\bibitem{ngo2014skew}
H.~Q. Ngo, C.~R{\'e}, and A.~Rudra.
\newblock Skew strikes back: new developments in the theory of join algorithms.
\newblock {\em ACM SIGMOD Record}, 42(4):5--16, 2014.

\bibitem{olteanu2016factorized}
D.~Olteanu and M.~Schleich.
\newblock Factorized databases.
\newblock {\em SIGMOD Record}, 45(2), 2016.

\bibitem{olteanu2012factorised}
D.~Olteanu and J.~Z{\'a}vodn{\`y}.
\newblock Factorised representations of query results: size bounds and
  readability.
\newblock In {\em ICDT}, 2012.

\bibitem{ordsDataset}
U.~M.~L. Repository.
\newblock Online retail data set.
\newblock \url{https://archive.ics.uci.edu/ml/datasets/Online+Retail}.

\bibitem{schleich2016learning}
M.~Schleich, D.~Olteanu, and R.~Ciucanu.
\newblock Learning linear regression models over factorized joins.
\newblock In {\em Proceedings of the 2016 International Conference on
  Management of Data}, pages 3--18. ACM, 2016.

\bibitem{schleich2019layered}
M.~Schleich, D.~Olteanu, and H.~Ngo.
\newblock A layered aggregate engine for analytics workloads.
\newblock 2019.

\bibitem{tarjan1984simple}
R.~E. Tarjan and M.~Yannakakis.
\newblock Simple linear-time algorithms to test chordality of graphs, test
  acyclicity of hypergraphs, and selectively reduce acyclic hypergraphs.
\newblock {\em SIAM Journal on computing}, 13(3):566--579, 1984.

\bibitem{veldhuizen2012leapfrog}
T.~L. Veldhuizen.
\newblock Leapfrog triejoin: A simple, worst-case optimal join algorithm.
\newblock {\em arXiv preprint arXiv:1210.0481}, 2012.

\bibitem{DBLP:conf/icdt/Veldhuizen14}
T.~L. Veldhuizen.
\newblock Triejoin: {A} simple, worst-case optimal join algorithm.
\newblock In {\em ICDT}, pages 96--106, 2014.

\bibitem{walenz2017optimizing}
B.~Walenz, S.~Roy, and J.~Yang.
\newblock Optimizing iceberg queries with complex joins.
\newblock In {\em SIGMOD}, 2017.

\bibitem{wang2010trie}
J.~Wang, J.~Feng, and G.~Li.
\newblock Trie-join: Efficient trie-based string similarity joins with
  edit-distance constraints.
\newblock {\em Proceedings of the VLDB Endowment}, 3(1-2):1219--1230, 2010.

\bibitem{xiao2011efficient}
C.~Xiao, W.~Wang, X.~Lin, J.~X. Yu, and G.~Wang.
\newblock Efficient similarity joins for near-duplicate detection.
\newblock {\em ACM Transactions on Database Systems (TODS)}, 36(3):15, 2011.

\bibitem{xin2003star}
D.~Xin, J.~Han, X.~Li, and B.~W. Wah.
\newblock Star-cubing: Computing iceberg cubes by top-down and bottom-up
  integration.
\newblock In {\em VLDB}, 2003.

\bibitem{yan1995interchanging}
W.~Yan and P.-A. Larson.
\newblock Interchanging the order of grouping and join.
\newblock Technical report, Technical Report CS 95-09, Dept. of Computer
  Science, University of Waterloo, Canada, 1995.

\bibitem{yan1994performing}
W.~P. Yan and P.-A. Larson.
\newblock Performing group-by before join.
\newblock In {\em ICDE}, 1994.

\bibitem{yan1995eager}
W.~P. Yan and P.-A. Larson.
\newblock Eager aggregation and lazy aggregation.
\newblock {\em Group}, 1:G2, 1995.

\end{thebibliography}

\end{document}